# Electrothermally Modulated Dispersion in Viscoelastic Fluids on Patterned Surfaces

Siddhartha Mukherjee,[1], Jayabrata Dhar[2], Sunando Dasgupta[1,3], Suman Chakraborty[1,2]

[1]*Advanced Technology Development Center, Indian Institute of Technology Kharagpur, Kharagpur, India-721302*
[2]*Department of Mechanical Engineering, Indian Institute of Technology Kharagpur, Kharagpur, India-721302*
[3]*Department of Chemical Engineering, Indian Institute of Technology Kharagpur, Kharagpur, India-721302*

Augmenting the dispersion of a solute species remains a challenging topic in electroosmotically-actuated flows due to the inherent plug-like feature of its velocity profile. In this study, we improvise on electrothermal flow of viscoelastic fluids through a parallel-plate microchannel with patterned surface charge to depict augmented dispersion coefficient of a solute. Here we consider the physical properties of the fluid to be temperature-dependent, thus, bringing out the combinatorial effect of both the thermal perturbation as well as the surface modulation on the resulting dynamics. This analysis unveils that by making a judicious choice of pertinent parameters one can control the flow rate with the desired account of dispersion. We envisage that the intricate coupling between the electrothermal perturbation and surface charge modulation combined with the fluid rheology may have tremendous implications towards the efficient designing of microfluidic devices.

\* E-mail address for correspondence: suman@mech.iitkgp.ernet.in

## I. INTRODUCTION

The miniaturization of microfluidic components has attracted significant attention in recent years owing to the emergence of "Lab on a chip" (LOC) devices [1,2] and subsequently the fluid flow within the narrow confinements is modulated with a target to achieve dedicated applications that primarily involve mixing [3–8] and separation. [9–11] Diffusion and dispersion are the two mechanisms which are highly advantageous for mixing purposes while for separation of chemical species they are undesirable. Hydrodynamic dispersion is a process in which band broadening of a neutral solute occurs because of the non-uniformity in the velocity distribution. [12,13] Because of its wide range of applications in several LOC processes, dispersion holds tremendous significance in the domain of microfluidics. [14–17] In recent years, non-Newtonian fluids has been employed in several micro and nanofluidic applications because of the close resemblance with biological fluids thus bearing enormous potentials in modern day medical and biotechnological research. [18–23] The behavior of these biofluids like blood plasma, saliva, synovial fluid, vitreous humor [24–28] exhibit viscoelasticity upon certain flow conditions, and hence, the use of viscoelastic models to predict their dynamics becomes prominent lately. [29–34]

Among various flow actuating mechanisms, electroosmosis stands as one of the most viable choices which finds several microfluidic uses ranging from medical diagnostics [35–38] to enhanced chip cooling. [39] In sharp contrast to pressure-driven flows, electroosmotic flow (EOF) typically exhibits plug-type velocity profile when the electrical double layer (EDL) thickness is small as compared to the channel dimension. Although more energy efficient, this uniformity in the velocity profile of EOF results in decline of solute dispersion in comparison with the pressure-driven flow. Thus, in order to improve the extent of mixing, the usual approach is to bring non-uniformity in the channel geometry or to alter the surface charge distribution, [40,41] thereby bringing out interesting flow visualizations with enhanced mixing [42,43]. Such surface charge modulation has applications in several fields like electrothermal flow, [44] thin film patterning, [45] AC electroosmosis, [46–48] induced-charge electroosmosis.[49,50] Consequently, the electrokinetically driven flow of complex viscoelastic fluids have also drawn special attention because of their huge applications ranging from species separation [51,52] to energy conversion in microfluidic devices. [53,54]

In addition, earlier reports of electro-hydrodynamic instabilities involving the interplay between the temperature gradients and concentration over small length scales may be exploited to further enhance solute dispersion in a medium. [55,56] Temperature gradients may occur either due to intrinsic Joule heating or via an external heat source. This developed non-isothermal condition results in the variation of the physical properties like viscosity, permittivity, electrical and thermal conductivity. For instance, the variation in the fluid conductivity interacting with the electric field can be used as a flow actuation mechanism, commonly termed as electro-convection flow. [56] This temperature dependence of the physical properties gives rise to drastic alteration in the flow field. Besides, the materials widely used for microchannel fabrication like polymethylmethacrylate (PMMA), polydimethylsiloxane (PDMS) have lower specific heat capacities and the lesser dissipation of heat from these materials may further contribute to augment the temperature gradients, thus, affecting the flow physics. [57] While most of the studies are directed towards the efficient transportation of fluid using the temperature-dependent property variation, the contribution of the electrothermal perturbation on the rotational behavior of the flow field remains relatively unexplored. More specifically, the combined effect of the electrothermal interaction and the fluid rheology along with the charge modulated surface condition, which may give rise to interesting flow controllability and augment dispersion of solutes, remains yet-to-be investigated, to the best of our knowledge.

Here we attempt to delineate the effect of electrothermal perturbations on the flow field and the associated dispersion coefficient in a patterned electrothermal flow of viscoelastic fluids through a parallel plate microchannel. The property variation coupled with the charge modulated surface significantly alters the flow field as well as the dispersion coefficient and it gets augmented when combined with the fluid rheology. We have developed an asymptotic approach following the classical lubrication approximation theory to take into account both the effect of electrothermal perturbation as well as the viscoelasticity on the dispersion coefficient. The results reported herein holds practical relevance in proper manipulation of flow in presence of finite temperature gradient, and thus, can be beneficial in the thermal management of microfluidic devices.

## II. PROBLEM FORMULATION

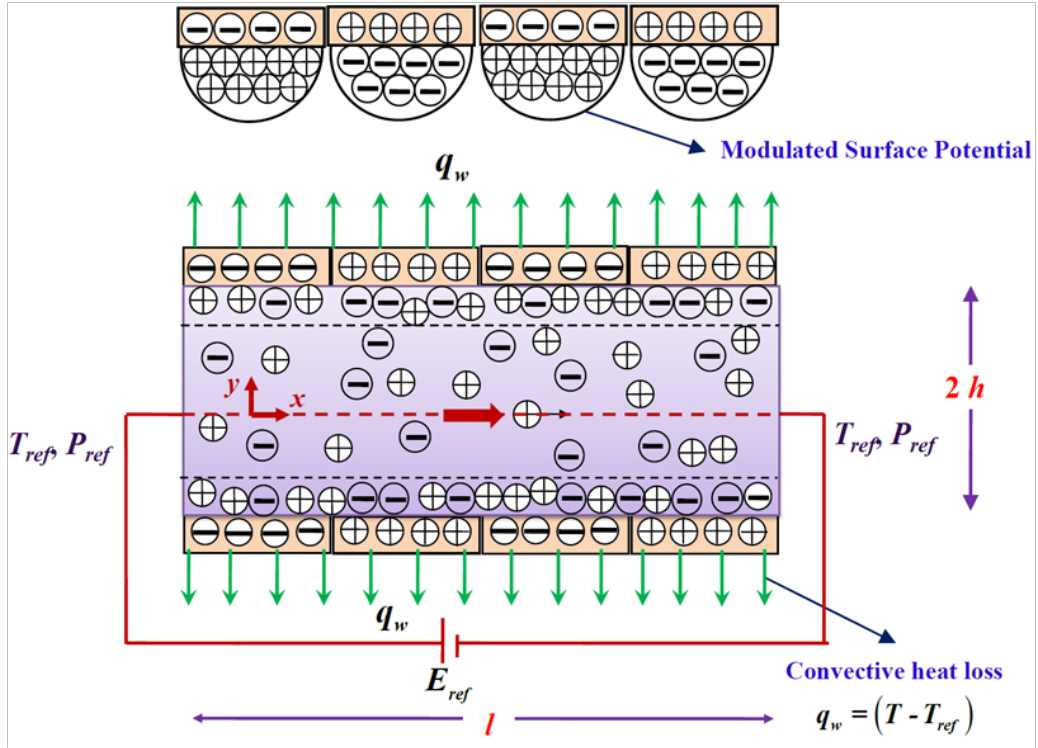

FIG. 1. Schematic of the patterned electrothermal flow in a parallel plate microchannel.

The representative diagram of patterned electrothermal flow of a viscoelastic fluid through a parallel plate microchannel is depicted in Fig. 1. The half-width of the microchannel is denoted by $h$, which is much smaller as compared to the length $(l)$ of the microchannel, i.e. $h \ll l$. The longitudinal and transverse coordinates are chosen along and perpendicular to the flow direction, respectively, with centerline at the inlet being the origin of the coordinate system, as depicted in the figure. For the patterned surface condition, we have chosen the following form of the zeta potential: $\zeta = \zeta_{ref}\{\alpha_1 + \alpha_2 \cos(\omega_t x)\}$, induced at the fluid-surface interface. Both ends of the microchannel are maintained at constant temperature $(T_{ref})$ and pressure $(p_{atm})$. An external electric field is considered to be applied in the axial direction. The heat generated due to the Joule heating effect is dissipated to the surrounding through natural convection. Before presenting the governing equations for the above system, it is necessary to discuss all assumptions required for this analysis.

In this study, we consider a symmetric binary electrolyte (*z:z*) being actuated by combined electrothermal and electroosmotic forces wherein the flow is assumed to be steady, laminar, incompressible and in the creeping flow regime, i.e. $Re \ll 1$. The physical properties of the fluid like viscosity $(\mu_{eff})$, thermal conductivity $(k_{eff})$, electrical conductivity $(\sigma_{eff})$, electrical permittivity $(\varepsilon_{eff})$ and relaxation time $(\lambda_{eff})$ are considered to be temperature dependent since these parameters are strongly influenced by temperature variation within the fluid domain. [58,59] We have assumed that the free ionic species within the EDL are point charges and they are in local equilibrium. Under this assumption, the advection term in the Poisson-Nernst-Plank equation can be safely neglected and the Poisson-Boltzmann description of the charge distribution remains valid, typically employed by several researchers. [40,41,60–62] (The detailed discussion for this consideration is included in **Appendix A**). For low values of induced electrical potential $(\zeta)$, (i.e. $e\zeta/k_B T \ll 1$ where $\zeta < 25$ mV), the well-known Debye-Hückel linearization approximation for the potential distribution can be applied. [41,43] Finally, we assume that the weak electric field approximation is valid, i.e. $E_{ref} \ll \zeta_{ref}/\lambda_D$ where $\lambda_D$ is the Debye length, very small as compared to the channel dimension $(h)$ thus precluding any distortion of the EDL structure. [41,63,64]

After taking into account the foregoing assumptions, the continuity, momentum and energy equations take the following form

$$\left.\begin{aligned}
\nabla \cdot \boldsymbol{v} &= 0 \\
\rho(\boldsymbol{v} \cdot \nabla)\boldsymbol{v} &= -\nabla p + \nabla \cdot \boldsymbol{\tau} + \boldsymbol{F}_b \\
\rho C_p (\boldsymbol{v} \cdot \nabla T) &= \nabla \cdot (k_{eff} \nabla T) + Q_{gen} + Q_{vd}
\end{aligned}\right\} \quad (1)$$

where $\boldsymbol{v}$ is the velocity vector, $\boldsymbol{\tau}$ is the stress tensor, $p$ is the hydrodynamic pressure and $T$ is the temperature field. $k_{eff}$ is the thermal conductivity of the fluid obeying the following relationship $k_{eff} = k_{ref}\left[1 + \alpha_3\left(T - T_{ref}\right)\right]$ [65,59] where $k_{ref}$ is the reference thermal conductivity evaluated at reference temperature $T_{ref}$. In the momentum equation, $F_b$ is the electrothermal body force defined by $F_b = \rho_e E - \frac{1}{2}|E|^2 \nabla \varepsilon_{eff} + \nabla\left\{\rho\left(\frac{\partial \varepsilon}{\partial \rho}\right)_T |E|^2\right\}$ [63,66] where $\rho_e$ is the free charge density and $E$ is the applied electric field written as $E = -\nabla \Phi$. The three terms in the body force are the Columbic force, dielectrophoretic force and electrostriction forces respectively. The Columbic force is induced due to the presence of free space charges and the variation in the electrical permittivity gives rise to the dielectrophoretic force while electrostriction force represents the compressibility of the fluid medium and thus can be neglected for incompressible fluids. In the energy equation, $Q_{gen}$ is the heat generated due to Joule heating effect, $Q_{gen} = \sigma_{eff}|E|^2$ and $Q_{vd}$ is the heat dissipation due to viscous action. $\sigma_{eff}$ and $\varepsilon_{eff}$ are the electrical conductivity and permittivities of the fluid which depend on temperature in the following way $\sigma_{eff} = \sigma_{ref}\left[1 + \alpha_4\left(T - T_{ref}\right)\right]$ and $\varepsilon_{eff} = \varepsilon_{ref}\left[1 - \alpha_5\left(T - T_{ref}\right)\right]$. [59,65] To determine the electric field, we employ the charge distribution equation along with the current continuity equation given by

$$\left.\begin{array}{l} \nabla \cdot \left(\varepsilon_{eff} \nabla \Phi\right) = -\rho_e \\ \nabla \cdot \boldsymbol{i} = \nabla \cdot (\sigma_{eff} \boldsymbol{E} + \boldsymbol{v}\, q - D \nabla q) = 0 \end{array}\right\} \quad (2)$$

The charge distribution subjected to the assumption of Poisson-Boltzmann description yields $\nabla \cdot \left(\varepsilon_{eff} \nabla \Phi\right) = -\rho_e = 2 n_0 z e \sinh\left(\frac{z e \Phi}{k_B T}\right)$ where $n_0$ is the number density of ions, $k_B$ Boltzmann constant, $z$ valence of ions, $e$ electronic charge and $T$ absolute temperature. In the current continuity equation, $q = -\rho_e$ and $D$ is the diffusion coefficient. Because of the thin EDL assumption, the contributions other than Ohmic current can be safely neglected. Thus, the electric potential becomes the sum of two potentials, one that is induced outside the EDL, another one that is induced within the EDL, i.e. $\Phi(x,y) = \phi(x) + \psi(x,y)$. Using this consideration, the Poisson-Boltzmann equation and the current continuity equation becomes

$$\left. \begin{array}{l} \dfrac{\partial}{\partial x}\left\{\varepsilon_{eff}\left(\dfrac{\partial \phi}{\partial x}+\dfrac{\partial \psi}{\partial x}\right)\right\}+\dfrac{\partial}{\partial y}\left\{\varepsilon_{eff}\left(\dfrac{\partial \psi}{\partial y}\right)\right\}=2n_0\,z\,e\sinh\left(\dfrac{z\,e\,\psi}{k_B T}\right)\\[2mm] \dfrac{\partial}{\partial x}\left\{\sigma_{eff}\left(\dfrac{\partial \phi}{\partial x}+\dfrac{\partial \psi}{\partial x}\right)\right\}+\dfrac{\partial}{\partial y}\left\{\sigma_{eff}\left(\dfrac{\partial \psi}{\partial y}\right)\right\}=0 \end{array} \right\} \qquad (3)$$

Employing the Debye-Hückel linearization, we have $\sinh\left(\dfrac{z\,e\,\psi}{k_B T}\right)\approx \dfrac{z\,e\,\psi}{k_B T}$. Thus the Poisson-Boltzmann equation recasts as $\nabla\cdot(\varepsilon_{eff}\nabla\Phi)=\dfrac{2n_0 z^2 e^2}{k_B T}\psi=\kappa_0^2\psi$ with $\kappa_0$ being the inverse of the EDL thickness defined as $\kappa_0=\sqrt{\dfrac{2n_0 z^2 e^2}{k_B T}}$. Now, we compare the contribution of the two components of the current continuity equation; $i_x\sim\sigma_{ref}E_{ref}$ and $i_y\sim\sigma_{ref}\zeta_{ref}/2h$ which gives $i_x/i_y\gg 1$ and the simplified equation is now integrated across the channel dimension

$$\left\{\int_{-h}^{h}\dfrac{\partial}{\partial x}\left\{\sigma\left(\dfrac{\partial \phi}{\partial x}+\dfrac{\partial \psi}{\partial x}\right)\right\}dy=0\right\} \qquad (4)$$

In the energy equation, one can neglect the relative contribution of the viscous dissipation term with respect to the other two terms since this effect becomes insignificant due to the thin EDL assumption

$$\rho C_p\left(u\dfrac{\partial T}{\partial x}+v\dfrac{\partial T}{\partial y}\right)=\dfrac{\partial}{\partial x}\left\{k_{eff}\dfrac{\partial T}{\partial x}\right\}+\dfrac{\partial}{\partial y}\left\{k_{eff}\dfrac{\partial T}{\partial y}\right\}+\sigma_{eff}\left\{\left(\dfrac{\partial \phi}{\partial x}+\dfrac{\partial \psi}{\partial x}\right)^2+\left(\dfrac{\partial \psi}{\partial y}\right)^2\right\} \qquad (5)$$

Here we introduce a new variable $\tilde{p}$ as $\tilde{p}=p-\dfrac{1}{2}\varepsilon_{ref}\kappa_0^2\psi^2$, [40,41] and rewrite the continuity and momentum equations in the following way

$$\left. \begin{array}{l} \dfrac{\partial u}{\partial x}+\dfrac{\partial v}{\partial y}=0\\[2mm] \rho\left(u\dfrac{\partial u}{\partial x}+v\dfrac{\partial u}{\partial y}\right)=-\dfrac{\partial \tilde{p}}{\partial x}+\dfrac{\partial \tau_{xx}}{\partial x}+\dfrac{\partial \tau_{yx}}{\partial y}+F_{bx}+\varepsilon_{eff}\kappa_0^2\psi\dfrac{\partial \psi}{\partial x}\\[2mm] \rho\left(u\dfrac{\partial v}{\partial x}+v\dfrac{\partial v}{\partial y}\right)=-\dfrac{\partial \tilde{p}}{\partial y}+\dfrac{\partial \tau_{xy}}{\partial x}+\dfrac{\partial \tau_{yy}}{\partial y}+F_{by}+\varepsilon_{eff}\kappa_0^2\psi\dfrac{\partial \psi}{\partial y} \end{array} \right\} \qquad (6)$$

In this study, we have considered the simplified Phan-Thien Tanner model (sPTT) [67,68] to simulate the viscoelastic fluid characteristics, typically used by several researchers. [29,31,32] Therefore, in the above equation, the stress components of the sPTT fluid, takes the form

$$
\begin{aligned}
2\mu_{eff} \frac{\partial u}{\partial x} &= F\,\tau_{xx} + \lambda_{eff}\left(u\frac{\partial \tau_{xx}}{\partial x} + v\frac{\partial \tau_{xx}}{\partial y} - 2\frac{\partial u}{\partial x}\tau_{xx} - 2\frac{\partial u}{\partial y}\tau_{yx}\right) \\
2\mu_{eff} \frac{\partial v}{\partial y} &= F\,\tau_{yy} + \lambda_{eff}\left(u\frac{\partial \tau_{yy}}{\partial x} + v\frac{\partial \tau_{yy}}{\partial y} - 2\frac{\partial v}{\partial x}\tau_{xy} - 2\frac{\partial v}{\partial y}\tau_{yy}\right) \\
\mu_{eff}\left(\frac{\partial u}{\partial y} + \frac{\partial v}{\partial x}\right) &= F\,\tau_{xy} + \lambda_{eff}\left(u\frac{\partial \tau_{xy}}{\partial x} + v\frac{\partial \tau_{xy}}{\partial y} - \frac{\partial u}{\partial y}\tau_{yy} - \frac{\partial v}{\partial x}\tau_{xx}\right)
\end{aligned}
\quad (7)
$$

where $F$ is the stress coefficient defined by $F = 1 + \dfrac{\delta\,\lambda_{eff}\left(\tau_{xx} + \tau_{yy}\right)}{\mu_{eff}}$ with $\delta$ representing the extensibility of the viscoelastic fluid. In Eq. (7), $\mu_{eff}$ and $\lambda_{eff}$ are the dynamic viscosity and relaxation time of the viscoelastic fluid which obey the following relationship: $\mu_{eff} = \mu_{ref}\exp\left[-\alpha_6\left(T - T_{ref}\right)\right]$ and $\lambda_{eff} = \lambda_{ref}\exp\left[-\alpha_6\left(T - T_{ref}\right)\right]$. [69,70] For the physical boundary conditions, we employ the classical no-slip condition at the channel walls while the $\zeta$-potentials follow the axially modulated profiles shown above. Isobaric and isothermal conditions are maintained at the two ends of the microchannel. The symmetry condition prevails at the channel centerline while heat loss occurs from the channel walls to the surrounding through natural convection, i.e. $q_w = h_T\left(\bar{T} - \bar{T}_{ref}\right)$ where $h_T$ is the convective heat transfer coefficient. Following this, the boundary conditions are represented in the following way

$$
\begin{aligned}
&u(y = \pm h) = 0,\ \ v(y = \pm h) = 0, \\
&p(x = 0) = p_{atm},\ \ p(x = l) = p_{atm}, \\
&T(x = 0) = T_{ref},\ \ T(x = 1) = T_{ref}, \\
&\phi(x = 0) = E_{ref}\,l = \phi_{ref},\ \ \phi(\bar{x} = 1) = 0, \\
&k_{eff}\left.\frac{\partial T}{\partial y}\right|_{y = \pm h} = \mp h_T\left(T - T_{ref}\right),\ \left.\frac{\partial T}{\partial y}\right|_{y=0} = 0, \\
&\psi(y = \pm h) = \zeta = \zeta_{ref}\left\{\alpha_1 + \alpha_2\cos(\omega_t x)\right\}
\end{aligned}
\quad (8)
$$

**Governing equations**

For solving the velocity, temperature and potential distribution, we non-dimensionalise the equation using the following dimensionless variables:

$$\left.\begin{array}{l} \bar{x}=\dfrac{x}{l},\ \bar{y}=\dfrac{y}{h},\ \bar{u}=\dfrac{u}{u_{HS}},\ \bar{v}=\dfrac{v\,l}{u_{HS}\,h},\ \bar{p}=\dfrac{(\tilde{p}-\tilde{p}_{atm})\,h^2}{\mu_{ref}\,u_{HS}\,l},\ \bar{T}=\dfrac{(T-T_{ref})}{\Delta T} \\[2mm] \bar{\tau}_{xx}=\dfrac{\tau_{xx}\,h}{\mu_{eff}\,u_{HS}},\ \bar{\tau}_{yy}=\dfrac{\tau_{yy}\,h}{\mu_{eff}\,u_{HS}},\ \bar{\tau}_{yx}=\dfrac{\tau_{yx}\,h}{\mu_{eff}\,u_{HS}},\ \bar{\phi}=\dfrac{\phi}{\phi_{ref}},\ \bar{\psi}=\dfrac{\psi}{\zeta_{ref}} \end{array}\right\} \quad (9)$$

where $u_{HS}=-\dfrac{\varepsilon_{ref}\,\zeta_{ref}\,E_{ref}}{\mu_{ref}}$ is the Helmholtz-Smoluchowski velocity, the characteristic scale for the electroosmotic flow and $\Delta T$ is the characteristic temperature difference defined as $\Delta T \sim \dfrac{\sigma_{ref}\,E_{ref}\,h\,l}{k_{ref}}$ obtained from equating the conduction term to the heat generation term in the energy equation. Using the aforementioned dimensionless variables, the continuity, momentum, energy and charge distribution equations are rewritten in the following way:

$$\text{Continuity Equation}: \quad \frac{\partial \bar{u}}{\partial \bar{x}}+\frac{\partial \bar{v}}{\partial \bar{y}}=0 \qquad (10)$$

$$\left.\begin{array}{l} x\text{-component}: \\[1mm] 0=-\dfrac{\partial \bar{p}}{\partial \bar{x}}+\chi\dfrac{\partial}{\partial \bar{x}}\left\{\exp(-\xi\bar{T})\bar{\tau}_{xx}\right\}+\dfrac{\partial}{\partial \bar{y}}\left\{\exp(-\xi\bar{T})\bar{\tau}_{yx}\right\}-\dfrac{\bar{\kappa}_o^2\,\bar{\psi}}{\xi\beta_1\bar{T}+1}\dfrac{\partial \bar{\phi}}{\partial \bar{x}} \\[3mm] -\bar{\kappa}_o^2\,\lambda\left\{\dfrac{\bar{\psi}}{\xi\beta_1\bar{T}+1}-\bar{\psi}\right\}\dfrac{\partial \bar{\psi}}{\partial \bar{x}}-\dfrac{\xi\beta_2}{2}\left\{\dfrac{\chi^2}{\lambda}\left(\dfrac{\partial \bar{\phi}}{\partial \bar{x}}+\lambda\dfrac{\partial \bar{\psi}}{\partial \bar{x}}\right)^2+\lambda\left(\dfrac{\partial \bar{\psi}}{\partial \bar{y}}\right)^2\right\}\dfrac{\partial \bar{T}}{\partial \bar{x}} \\[3mm] y\text{-component}: \\[1mm] 0=-\dfrac{\partial \bar{p}}{\partial \bar{y}}+\chi^2\dfrac{\partial}{\partial \bar{x}}\left\{\exp(-\xi\bar{T})\bar{\tau}_{xy}\right\}+\chi\dfrac{\partial}{\partial \bar{y}}\left\{\exp(-\xi\bar{T})\bar{\tau}_{yy}\right\} \\[3mm] -\bar{\kappa}_o^2\lambda\left\{\left(\dfrac{\bar{\psi}}{\xi\beta_1\bar{T}+1}\right)-\bar{\psi}\right\}\dfrac{\partial \bar{\psi}}{\partial \bar{y}}-\dfrac{\xi\beta_2}{2}\left\{\dfrac{\chi^2}{\lambda}\left(\dfrac{\partial \bar{\phi}}{\partial \bar{x}}+\lambda\dfrac{\partial \bar{\psi}}{\partial \bar{x}}\right)^2+\lambda\left(\dfrac{\partial \bar{\psi}}{\partial \bar{y}}\right)^2\right\}\dfrac{\partial \bar{T}}{\partial \bar{y}} \end{array}\right\} \quad (11)$$

Energy Equation:

$$Pe_T\left(\bar{u}\frac{\partial \bar{T}}{\partial \bar{x}}+v\frac{\partial \bar{T}}{\partial \bar{y}}\right)=\chi\frac{\partial}{\partial \bar{x}}\left\{(1+\xi\beta_3\bar{T})\frac{\partial \bar{T}}{\partial \bar{x}}\right\}+\frac{1}{\chi}\frac{\partial}{\partial \bar{y}}\left\{(1+\xi\beta_3\bar{T})\frac{\partial \bar{T}}{\partial \bar{y}}\right\} \quad (12)$$

$$+(1+\xi\beta_4\bar{T})\left\{\left(\frac{\partial \bar{\phi}}{\partial \bar{x}}\right)^2+2\lambda\frac{\partial \bar{\phi}}{\partial \bar{x}}\frac{\partial \bar{\psi}}{\partial \bar{x}}+\lambda^2\left(\frac{\partial \bar{\psi}}{\partial \bar{x}}\right)^2\right\}+\frac{\lambda^2(1+\xi\beta_4\bar{T})}{\chi^2}\left(\frac{\partial \bar{\psi}}{\partial \bar{y}}\right)^2$$

In Eqs. (11)-(12), $\chi$ is the lubrication parameter $(\chi=h/l)$ and $Pe_T$ is the thermal Peclet number $(Pe_T=\rho C_p u_{HS} h/k_{ref})$. Here, $\lambda=\dfrac{\zeta_{ref}}{\phi_{ref}}$ is the ratio of the induced potential and the applied potential and $\beta_1=\dfrac{1}{\alpha_6 T_{ref}}$, $\beta_2=\dfrac{\alpha_5}{\alpha_6}$, $\beta_3=\dfrac{\alpha_3}{\alpha_6}$, $\beta_4=\dfrac{\alpha_4}{\alpha_6}$ are the parameters showing the temperature dependence of the physical properties while $\bar{\kappa}_0=\kappa_0 h$ is the inverse of the dimensionless EDL thickness. Besides, the thermal perturbation to the system can be characterized by the parameter $\xi=\alpha_6\Delta T$. Also, we define a new variable $v=\dfrac{h_T h}{k_{ref}}$ to take into account the convective heat loss to the surrounding. Now, considering the case of low surface potential, the simplified charge distribution along with the current continuity equations are presented below

Charge Distribution:
$$\frac{\bar{\kappa}_0^2\bar{\psi}}{\xi\beta_1\bar{T}+1}=\frac{\chi^2}{\lambda}\frac{\partial}{\partial \bar{x}}\left\{(1-\xi\beta_2\bar{T})\left(\frac{\partial \bar{\phi}}{\partial \bar{x}}+\lambda\frac{\partial \bar{\psi}}{\partial \bar{x}}\right)\right\}+\frac{\partial}{\partial \bar{y}}\left\{(1-\xi\beta_2\bar{T})\frac{\partial \bar{\psi}}{\partial \bar{y}}\right\}$$

Current Continuity:
$$\int_{-1}^{1}\frac{\partial}{\partial \bar{x}}\left\{(1+\xi\beta_4\bar{T})\left(\frac{\partial \bar{\phi}}{\partial \bar{x}}+\lambda\frac{\partial \bar{\psi}}{\partial \bar{x}}\right)\right\}d\bar{y}=0 \quad (13)$$

After using the relevant scales, the dimensionless forms of the stress components for a viscoelastic fluid takes the following form

$$2\chi\frac{\partial \bar{u}}{\partial \bar{x}} = \left\{1 + \frac{\delta\, De_{\bar{\kappa}_o}}{\bar{\kappa}_0}\exp(-\xi\,\bar{T})(\bar{\tau}_{xx}+\bar{\tau}_{yy})\right\}\bar{\tau}_{xx} + \frac{De_{\bar{\kappa}_o}\chi}{\bar{\kappa}_0}\left\{\begin{array}{l}\bar{u}\dfrac{\partial}{\partial \bar{x}}\{\exp(-\xi\,\bar{T})\bar{\tau}_{xx}\} + \bar{v}\dfrac{\partial}{\partial \bar{y}}\{\exp(-\xi\,\bar{T})\bar{\tau}_{xx}\} \\ -2\exp(-\xi\,\bar{T})\dfrac{\partial \bar{u}}{\partial \bar{x}}\bar{\tau}_{xx} - \dfrac{2}{\chi}\exp(-\xi\,\bar{T})\dfrac{\partial \bar{u}}{\partial \bar{y}}\bar{\tau}_{yx}\end{array}\right\}$$

$$2\chi\frac{\partial \bar{v}}{\partial \bar{y}} = \left\{1 + \frac{\delta\, De_{\bar{\kappa}_o}}{\bar{\kappa}_0}\exp(-\xi\,\bar{T})(\bar{\tau}_{xx}+\bar{\tau}_{yy})\right\}\bar{\tau}_{yy} + \frac{De_{\bar{\kappa}_o}\chi}{\bar{\kappa}_0}\left\{\begin{array}{l}\bar{u}\dfrac{\partial}{\partial \bar{x}}\{\exp(-\xi\,\bar{T})\bar{\tau}_{yy}\} + \bar{v}\dfrac{\partial}{\partial \bar{y}}\{\exp(-\xi\,\bar{T})\bar{\tau}_{yy}\} \\ -2\chi\exp(-\xi\,\bar{T})\dfrac{\partial \bar{v}}{\partial \bar{x}}\bar{\tau}_{xy} - 2\exp(-\xi\,\bar{T})\dfrac{\partial \bar{v}}{\partial \bar{y}}\bar{\tau}_{yy}\end{array}\right\} \quad (14)$$

$$\frac{\partial \bar{u}}{\partial \bar{y}} + \chi^2\frac{\partial \bar{v}}{\partial \bar{x}} = \left\{1 + \frac{\delta\, De_{\bar{\kappa}_o}}{\bar{\kappa}_0}\exp(-\xi\,\bar{T})(\bar{\tau}_{xx}+\bar{\tau}_{yy})\right\}\bar{\tau}_{xy} + \frac{De_{\bar{\kappa}_o}\chi}{\bar{\kappa}_0}\left\{\begin{array}{l}\bar{u}\dfrac{\partial}{\partial \bar{x}}\{\exp(-\xi\bar{T})\bar{\tau}_{xy}\} + \bar{v}\dfrac{\partial}{\partial \bar{y}}\{\exp(-\xi\bar{T})\bar{\tau}_{xy}\} \\ -\chi\exp(-\xi\bar{T})\dfrac{\partial \bar{v}}{\partial \bar{x}}\bar{\tau}_{xx} - \dfrac{\exp(-\xi\bar{T})}{\chi}\dfrac{\partial \bar{u}}{\partial \bar{y}}\bar{\tau}_{yy}\end{array}\right\}$$

where Deborah number ($De_{\bar{\kappa}_0} = \lambda_{ref}\kappa_0 u_{HS}$) represents the extent of viscoelasticity of the fluid wherein $De_{\bar{\kappa}_0} = 0$ corresponds to Newtonian fluid. Now, the boundary conditions described by Eq. (8) are rewritten in their respective non-dimensional forms

$$\left.\begin{array}{l}\bar{u}(\bar{y}=\pm 1)=0,\ \bar{v}(\bar{y}=\pm 1)=0, \\ \bar{p}(\bar{x}=0)=0,\ \bar{p}(\bar{x}=1)=0, \\ \bar{T}(x=0)=0,\ \bar{T}(\bar{x}=1)=0, \\ \bar{\phi}(x=0)=1,\ \bar{\phi}(\bar{x}=1)=0, \\ (1+\xi\beta_3\bar{T})\left.\dfrac{\partial \bar{T}}{\partial \bar{y}}\right|_{\bar{y}=\pm 1} = \mp v\,\bar{T},\ \left.\dfrac{\partial \bar{T}}{\partial \bar{y}}\right|_{\bar{y}=0}=0, \\ \bar{\psi}(\bar{y}=\pm 1)=\alpha_1 + \alpha_2\cos(\omega\,\bar{x})\end{array}\right\} \quad (15)$$

where $\omega = \omega_t l$ is the patterning frequency of modulation. To obtain the flow and temperature fields from the set of above dimensionless forms, we have performed an asymptotic approach followed by the classical lubrication approximation theory. [71–73] In typical microfluidic applications, the length scale in the transverse coordinate is very small as compared to the longitudinal co-ordinates $(l \gg h)$. In the limit of $\chi \to 0$, the terms involving $\mathrm{O}(\chi)$ and

$O(\chi^2)$ can be discarded and the simplified momentum and the stress components take the following form

$$\begin{aligned}
\frac{\partial \overline{p}}{\partial \overline{x}} &= \frac{\partial}{\partial \overline{y}}\left\{\exp(-\xi \overline{T})\overline{\tau}_{yx}\right\} - \frac{\overline{\kappa}_o^2 \overline{\psi}}{\xi \beta_1 \overline{T}+1}\frac{\partial \overline{\phi}}{\partial \overline{x}} - \overline{\kappa}_o^2 \lambda \left\{\frac{\overline{\psi}}{\xi \beta_1 \overline{T}+1} - \overline{\psi}\right\}\frac{\partial \overline{\psi}}{\partial \overline{x}} \\
&\quad - \frac{\xi \beta_2}{2}\left\{\frac{\chi^2}{\lambda}\left(\frac{\partial \overline{\phi}}{\partial \overline{x}}+\lambda\frac{\partial \overline{\psi}}{\partial \overline{x}}\right)^2 + \lambda\left(\frac{\partial \overline{\psi}}{\partial \overline{y}}\right)^2\right\}\frac{\partial \overline{T}}{\partial \overline{x}} \\
\frac{\partial \overline{p}}{\partial \overline{y}} &= -\overline{\kappa}_o^2 \lambda\left\{\frac{\overline{\psi}}{\xi \beta_1 \overline{T}+1}-\overline{\psi}\right\}\frac{\partial \overline{\psi}}{\partial \overline{y}} - \frac{\xi \beta_2}{2}\left\{\frac{\chi^2}{\lambda}\left(\frac{\partial \overline{\phi}}{\partial \overline{x}}+\lambda\frac{\partial \overline{\psi}}{\partial \overline{x}}\right)^2 + \lambda\left(\frac{\partial \overline{\psi}}{\partial \overline{y}}\right)^2\right\}\frac{\partial \overline{T}}{\partial \overline{y}}
\end{aligned} \quad (16)$$

$$\begin{aligned}
\left\{1+\frac{\delta \, De_{\overline{\kappa}_o}}{\overline{\kappa}_0}\exp(-\xi \overline{T})(\overline{\tau}_{xx}+\overline{\tau}_{yy})\right\}\overline{\tau}_{xx} &= \frac{2\,De_{\overline{\kappa}_o}}{\overline{\kappa}_0}\exp(-\xi \overline{T})\frac{\partial \overline{u}}{\partial \overline{y}}\overline{\tau}_{yx} \\
\left\{1+\frac{\delta \, De_{\overline{\kappa}_o}}{\overline{\kappa}_0}\exp(-\xi \overline{T})(\overline{\tau}_{xx}+\overline{\tau}_{yy})\right\}\overline{\tau}_{yy} &= 0 \\
\frac{\partial \overline{u}}{\partial \overline{y}} + \frac{De_{\overline{\kappa}_o}}{\overline{\kappa}_0}\exp(-\xi \overline{T})\frac{\partial \overline{u}}{\partial \overline{y}}\overline{\tau}_{yy} &= \left\{1+\frac{\delta \, De_{\overline{\kappa}_o}}{\overline{\kappa}_0}\exp(-\xi \overline{T})(\overline{\tau}_{xx}+\overline{\tau}_{yy})\right\}\overline{\tau}_{xy}
\end{aligned} \quad (17)$$

From Eq. (17), it is clear that $\overline{\tau}_{yy}=0$ and a relationship between the stress components $\overline{\tau}_{xx}$ and $\overline{\tau}_{yx}$ can be established. Meanwhile in the energy equation, the conduction terms cannot be neglected even in the limit of $\chi \to 0$ because of their relative strengths with respect to the other terms. To determine the relative contributions of the convective components, we compare the respective scales of the characteristic temperature differences which are $\Delta T_x \sim \dfrac{\sigma_{ref} E_{ref} l^2}{k_{ref}}$ and $\Delta T_y \sim \dfrac{\sigma_{ref} E_{ref} h^2}{k_{ref}}$ respectively and hence, $\dfrac{\Delta T_x}{\Delta T_y} \sim \dfrac{l^2}{h^2} \gg 1$, i.e., $\dfrac{\partial \overline{T}}{\partial \overline{y}} \ll \dfrac{\partial \overline{T}}{\partial \overline{x}}$. Additionally, one can assume that the surface potential is very small as compared to the applied potential, i.e., $\dfrac{\zeta_{ref}}{\phi_{ref}} = \lambda \ll 1$ and the terms involving $O(\lambda)$ and its higher orders can be neglected thereby resulting the following simplified forms

$$\left. \begin{array}{l} \dfrac{\partial \overline{p}}{\partial \overline{x}} = \dfrac{\partial}{\partial \overline{y}}\left\{\exp\left(-\xi \overline{T}\right)\overline{\tau}_{yx}\right\} - \dfrac{\overline{\kappa}_0^2 \overline{\psi}}{\xi \beta_1 \overline{T}+1}\dfrac{\partial \overline{\phi}}{\partial \overline{x}} - \dfrac{\xi \beta_2 \chi^2}{2\lambda}\left(\dfrac{\partial \overline{\phi}}{\partial \overline{x}}\right)^2 \dfrac{\partial \overline{T}}{\partial \overline{x}} \\[2ex] \dfrac{\partial \overline{p}}{\partial \overline{y}} = -\dfrac{\xi \beta_2 \chi^2}{2\lambda}\left(\dfrac{\partial \overline{\phi}}{\partial \overline{x}}\right)^2 \dfrac{\partial \overline{T}}{\partial \overline{y}} \end{array} \right\} \quad (18)$$

$$Pe_T\left(\overline{u}\dfrac{\partial \overline{T}}{\partial \overline{x}}\right) = \chi\dfrac{\partial}{\partial \overline{x}}\left\{(1+\xi \beta_3 \overline{T})\dfrac{\partial \overline{T}}{\partial \overline{x}}\right\} + \dfrac{1}{\chi}\dfrac{\partial}{\partial \overline{y}}\left\{(1+\xi \beta_3 \overline{T})\dfrac{\partial \overline{T}}{\partial \overline{y}}\right\} + (1+\xi \beta_4 \overline{T})\left(\dfrac{\partial \overline{\phi}}{\partial \overline{x}}\right)^2 \quad (19)$$

Now we compare the relative contributions of the terms of the charge distribution described by Eq. (3). Choosing appropriate scales of the respective parameters, i.e. $\varepsilon \sim \varepsilon_{ref}$, $\phi \sim \phi_{ref}$, $\psi \sim \zeta_{ref}$, $x \sim l$, and $y \sim \lambda_D$, the first term on the left hand side becomes $\sim \dfrac{\varepsilon_{ref}}{l^2}(\phi_{ref}+\zeta_{ref})$ while the second term is scaled as $\sim \dfrac{\varepsilon_{ref}\zeta_{ref}}{\lambda_D^2}$. Hence, the ratio becomes $\sim \dfrac{\varepsilon_{ref}}{l^2}(\phi_{ref}+\zeta_{ref})\bigg/\dfrac{\varepsilon_{ref}\zeta_{ref}}{\lambda_D^2}$

$\sim \dfrac{\lambda_D^2}{l^2} + \dfrac{\lambda_D^2}{l^2\lambda} \sim \dfrac{\lambda_D^2}{l^2} + \dfrac{\chi^2}{\kappa_0^2\lambda}$ where $\lambda_D \ll l$ and $\dfrac{\chi^2}{\kappa_0^2\lambda} \ll 1$; which is negligible compared to the second term and the simplified form is now written below

$$\left. \begin{array}{l} \dfrac{\overline{\kappa}_0^2 \overline{\psi}}{\xi \beta_1 \overline{T}+1} = \dfrac{\partial}{\partial \overline{y}}\left\{(1-\xi \beta_2 \overline{T})\dfrac{\partial \overline{\psi}}{\partial \overline{y}}\right\} \\[2ex] \text{and} \quad \int_{-1}^{1}\dfrac{\partial}{\partial \overline{x}}\left\{(1+\xi \beta_4 \overline{T})\left(\dfrac{\partial \overline{\phi}}{\partial \overline{x}}\right)\right\}d\overline{y} = 0 \end{array} \right\} \quad (20)$$

By choosing the typical values of the pertinent parameters (these values are shown in the Results and Discussion section), one can show that $\dfrac{\xi \beta_2 \chi^2}{2\lambda} \ll 1$ and the momentum components are then reduced to the following form

$$\left. \begin{array}{l} \dfrac{d\overline{p}}{d\overline{x}} = \dfrac{\partial}{\partial \overline{y}}\left\{\exp\left(-\xi \overline{T}\right)\overline{\tau}_{yx}\right\} - \dfrac{\overline{\kappa}_0^2 \overline{\psi}}{\xi \beta_1 \overline{T}+1}\dfrac{\partial \overline{\phi}}{\partial \overline{x}} \\[2ex] \dfrac{\partial \overline{p}}{\partial \overline{y}} = 0 \end{array} \right\} \quad (21)$$

Since the axial variation of the temperature in the *x* co-ordinate is more significant compared to the *y* co-ordinate, one can expand the temperature distribution in an asymptotic series in the following manner

$$\bar{T} = \bar{T}_0(\bar{x}) + \nu\, \bar{T}_1(\bar{x},\bar{y}) + \nu^2\, \bar{T}_2(\bar{x},\bar{y}) + O(\nu^3) \tag{22}$$

where $\nu$ characterizes the rate of heat loss to the surrounding. Now, we utilize this expansion along with the two thermal boundary conditions and integrate the energy equation over the entire domain

$$\frac{1}{2} Pe \frac{d\bar{T}_0}{d\bar{x}} \left\{ \int_{-1}^{1} \bar{u}\, d\bar{y} \right\} = \chi \frac{d^2 \bar{T}_0}{d\bar{x}^2} - \frac{\nu\, \bar{T}_0}{\chi} + \left(1 + \xi \beta_4 \bar{T}_0\right) \left(\frac{d\bar{\phi}}{d\bar{x}}\right)^2 \tag{23}$$

Now, the potential distribution for the patterned electrothermal flow is given by

$$\bar{\psi} = \{\alpha_1 + \alpha_2 \cos(\omega \bar{x})\} \frac{\cosh\left\{ \left( \frac{\bar{\kappa}_0^2}{1 + (\beta_1 - \beta_2)\xi \bar{T}_0 - \beta_1 \beta_2 \xi^2 \bar{T}_0} \right)^{1/2} \bar{y} \right\}}{\cosh\left( \frac{\bar{\kappa}_0^2}{1 + (\beta_1 - \beta_2)\xi \bar{T}_0 - \beta_1 \beta_2 \xi^2 \bar{T}_0} \right)^{1/2}} \tag{24}$$

Using typical values of the involving parameters, one can show that $(\beta_1 - \beta_2)\xi \ll 1$ and $\beta_1 \beta_2 \xi^2 \ll 1$. Hence, the potential distribution is simplified and takes the following form

$$\bar{\psi} = \{\alpha_1 + \alpha_2 \cos(\omega \bar{x})\} \frac{\cosh\{\bar{\kappa}_0\, \bar{y}\}}{\cosh(\bar{\kappa}_0)} \tag{25}$$

Using the expansion of Eq. (22), the set of governing equations are rewritten below

$$\left. \begin{aligned} & \frac{\partial \bar{u}}{\partial \bar{x}} + \frac{\partial \bar{v}}{\partial \bar{y}} = 0 \\ & \frac{d\bar{p}}{d\bar{x}} = \frac{\partial}{\partial \bar{y}}\left\{ \exp(-\xi \bar{T}_0) \bar{\tau}_{yx} \right\} - \bar{\kappa}_o^2 \{\alpha_1 + \alpha_2 \cos(\omega \bar{x})\} \frac{\cosh(\bar{\kappa}_0\, \bar{y})}{\cosh(\bar{\kappa}_0)} \frac{\partial \bar{\phi}}{\partial \bar{x}} \\ & \frac{1}{2} Pe \frac{d\bar{T}_0}{d\bar{x}} \left\{ \int_{-1}^{1} \bar{u}\, d\bar{y} \right\} = \chi \frac{d^2 \bar{T}_0}{d\bar{x}^2} - \frac{\nu \bar{T}_0}{\chi} + \left(1 + \xi \beta_4 \bar{T}_0\right) \left(\frac{d\bar{\phi}}{d\bar{x}}\right)^2 \\ & \text{and} \qquad \frac{\partial}{\partial \bar{x}}\left\{ \left(1 + \xi \beta_4 \bar{T}_0\right) \frac{\partial \bar{\phi}}{\partial \bar{x}} \right\} = 0 \end{aligned} \right\} \tag{26}$$

The simplified stress components (after substituting $\bar{\tau}_{yy} = 0$) are also expanded in a similar way

$$\frac{\partial \bar{u}}{\partial \bar{y}} = \left\{ 1 + \frac{2\delta\, De_{\bar{\kappa}_o}^2}{\bar{\kappa}_0^2} \exp(-2\xi \bar{T}_0)(\bar{\tau}_{yx}) \right\} \bar{\tau}_{yx} \tag{27}$$

In order to solve the Eqs. (26)-(27), we have used an asymptotic approach which is described in detail in **Appendix B**.

## LIMITING CASES

On the basis of the present asymptotic analysis, we investigate some limiting cases.

**Case 1:** If we substitute $\xi = 0$, the velocity profile reduces to the following form

$$\bar{u} = \frac{1}{2}\frac{d\bar{p}_{0,0}}{d\bar{x}}(\bar{y}^2 - 1) + \{\alpha_1 + \alpha_2 \cos(\omega\bar{x})\}\left[1 - \frac{\cosh(\bar{\kappa}_0 \bar{y})}{\cosh(\bar{\kappa}_0)}\right] + \frac{1}{2}\frac{d\bar{p}_{0,1}}{d\bar{x}}(\bar{y}^2 - 1)$$

$$+ \frac{2\delta De^2}{\bar{\kappa}_0^2}\begin{bmatrix} \frac{a_1^3}{4}\left\{\cos(\omega\bar{x}) - \frac{\sin(\omega)}{\omega}\right\}^3 (\bar{y}^4 - 1) - \frac{\bar{\kappa}_0^2}{12}\frac{\{\alpha_1 + \alpha_2 \cos(\omega\bar{x})\}^3}{\cosh^3(\bar{\kappa}_0)} f_3(\bar{y}) \\ -\frac{3a_1^2\{\alpha_1 + \alpha_2 \cos(\omega\bar{x})\}}{\bar{\kappa}_0^2 \cosh(\bar{\kappa}_0)}\left\{\cos(\omega\bar{x}) - \frac{\sin(\omega)}{\omega}\right\}^2 f_1(\bar{y}) \\ +\frac{3a_1\{\alpha_1 + \alpha_2 \cos(\omega\bar{x})\}^2}{8\cosh^2(\bar{\kappa}_0)}\left\{\cos(\omega\bar{x}) - \frac{\sin(\omega)}{\omega}\right\} f_2(\bar{y}) \end{bmatrix} \quad (28)$$

which represents the flow field for a patterned electroosmotic flow of a viscoelastic fluid in absence of any thermal perturbation with the coefficients given in **Appendix C**. Further simplification is possible by substituting $\alpha_2 = 0$, $\alpha_1 = 1$ which results

$$\bar{u} = \left[1 - \frac{\cosh(\bar{\kappa}_0 \bar{y})}{\cosh(\bar{\kappa}_0)}\right] - \frac{\delta De^2}{6\cosh^3(\bar{\kappa}_0)}\left[\cosh(3\bar{\kappa}_0 \bar{y}) - 9\cosh(\bar{\kappa}_0 \bar{y}) - \cosh(3\bar{\kappa}_0) + 9\cosh(\bar{\kappa}_0)\right] \quad (29)$$

which is the velocity profile for purely electroosmotic flow of viscoelastic fluid through parallel plate microchannel. [29]

**Case 2:** If we substitute $De^* = 0$, the velocity profile reduces to the following form

$$\bar{u} = \frac{1}{2}\frac{d\bar{p}_{0,0}}{d\bar{x}}(\bar{y}^2 - 1) + \{\alpha_1 + \alpha_2 \cos(\omega\bar{x})\}\left[1 - \frac{\cosh(\bar{\kappa}_0 \bar{y})}{\cosh(\bar{\kappa}_0)}\right]$$

$$+ \xi \begin{bmatrix} \frac{1}{2}\frac{d\bar{p}_{1,0}}{d\bar{x}}(\bar{y}^2 - 1) + \frac{1}{2}\bar{T}_{0,0,0}\frac{d\bar{p}_{0,0}}{d\bar{x}}(\bar{y}^2 - 1) + \{\alpha_1 + \alpha_2 \cos(\omega\bar{x})\}\left\{\frac{\cosh(\bar{\kappa}_0 \bar{y})}{\cosh(\bar{\kappa}_0)} - 1\right\}\frac{d\bar{\phi}_{1,0}}{d\bar{x}} \\ +\bar{T}_{0,0,0}\{\alpha_1 + \alpha_2 \cos(\omega\bar{x})\}\left\{1 - \frac{\cosh(\bar{\kappa}_0 \bar{y})}{\cosh(\bar{\kappa}_0)}\right\} \end{bmatrix} \quad (30)$$

which is the velocity distribution for patterned electrothermal flow of a Newtonian fluid which is further reduced on substitution of $\alpha_2 = 0$, $\alpha_1 = 1$

$$\overline{u} = \left[1 - \frac{\cosh(\overline{\kappa}_0 \overline{y})}{\cosh(\overline{\kappa}_0)}\right] + \xi\left[\left\{\frac{\cosh(\overline{\kappa}_0 \overline{y})}{\cosh(\overline{\kappa}_0)} - 1\right\}\frac{d\overline{\phi}_{1,0}}{d\overline{x}} + \overline{T}_{0,0,0}\left\{1 - \frac{\cosh(\overline{\kappa}_0 \overline{y})}{\cosh(\overline{\kappa}_0)}\right\}\right] \quad (31)$$

The expressions for $\overline{T}_{0,0,0}$ and $\dfrac{d\overline{\phi}_{1,0}}{d\overline{x}}$ can be found in **Appendix F**.

**Stream function**

For better visualization of the flow structure, we use the definition of stream function and express them in their non-dimensional form as $\overline{u} = -\dfrac{\partial \overline{\varphi}}{\partial \overline{y}}$ and $\overline{v} = \dfrac{\partial \overline{\varphi}}{\partial \overline{x}}$ where $\overline{\varphi} = \dfrac{\varphi}{h\, u_{HS}}$. To address the effect of thermal perturbation on the flow field, we expand $\overline{\varphi}$ in the asymptotic series as

$$\overline{\varphi} = \overline{\varphi}_0 + \xi\,\overline{\varphi}_1 + \xi^2\,\overline{\varphi}_2 + \cdots \quad (32)$$

which is subjected to the following boundary condition

$$\overline{\varphi} = 0 \quad \text{at} \quad \overline{y} = \pm 1$$

Since, the expression of the stream function is large; it is not included in the Appendix section for the sake of conciseness. The MATLAB scripts containing detailed expressions can be made available upon request.

**Dispersion coefficient**

We have considered the dispersion occurring owing to an interaction between the electric field and flow field combined with the charge modulated surface. From definition, the dispersion coefficient $(D_{eff})$ is related to the solute distribution in the following way [73]

$$D_{eff} = \frac{1}{2}\frac{d}{dt}\sigma^2(t) \quad (33)$$

where $\sigma^2$ is the variance in the solute displacement band which in turn is related to the plate height $(\tilde{h})$ as

$$\tilde{h} = \frac{d}{d\tilde{x}}\sigma^2(\tilde{x}) \quad (34)$$

with $\tilde{x}$ being the centre of mass of the band. Thus, $\sigma^2$ and $\tilde{x}$ are the two parameters which are associated with the band broadening phenomenon where the velocity of the centre of mass is given by

$$\tilde{u} = \frac{d\tilde{x}}{dt} \tag{35}$$

Now, combining these two equations one can rewrite the dispersion coefficient $\left(D_{eff}\right)$ as

$$D_{eff} = \frac{1}{2}\tilde{u}\tilde{h} \tag{36}$$

where $\tilde{u}$ also represents the cross-sectional averaged flow velocity through the microchannel and the plate height $\tilde{h}$ is evaluated by

$$\tilde{h} = \frac{2D}{\tilde{u}} + \frac{\tilde{u}}{8D}h^{*2} \tag{37}$$

where $D$ is the molecular coefficient and $h^*$ is the minimum plate height which is given by [73]

$$h^* = \frac{16}{h}\int_0^h \int_0^y \left\{\left(\frac{u}{\tilde{u}}-1\right)dy\right\}^2 dy \tag{38}$$

Now, the dimensionless form of Eq. (36) becomes

$$\bar{D}_{eff} = 1 + \frac{1}{16}\left(Pe_D \bar{\tilde{u}} h^*\right)^2 \tag{39}$$

where $\bar{D}_{eff}$ is the dimensionless dispersion coefficient, $Pe_D$ is the Peclet number for dispersion and $\bar{\tilde{u}}$ is the dimensionless average velocity.

## III. RESULTS AND DISCUSSIONS

For presenting the results of the velocity, temperature and potential fields, we have chosen some typical values of the pertinent parameters: $h \sim 10-10^2$ μm, $l \sim 1-10$ mm, $\rho \sim 10^3 \text{ kg/m}^3$, $C_p = 4200 \text{ J/kg}\cdot\text{K}$, $E_{ref} \sim 10^3-10^4 \text{ V/m}$, $\zeta_{ref} \sim 10^{-2}\text{V}$, $\mu_{ref} \sim 10^{-3}\text{Pa.s}$, $\sigma_{ref} \sim 10^{-2}\text{-}10^{-1} \text{ S/m}$, $k_{ref} \sim 0.613 \text{ W/m}\cdot\text{K}$, $\varepsilon_{ref} \sim 10^{-10} \text{ CV}^{-1}\text{m}^{-1}$, $\lambda_{ref} \sim 10^{-3}-10^{-1}$ s, $T_{ref} \sim 298$ K, $D \sim 10^{-9} \text{ m}^2\text{s}^{-1}$, $\Delta T \sim 1-10^2$ K, $\lambda_D \sim 1-100$ nm, $\alpha_3 \sim 10^{-3} \text{ K}^{-1}$, $\alpha_4 \sim 10^{-2} \text{ K}^{-1}$, $\alpha_5 \sim 10^{-3} \text{ K}^{-1}$ and $\alpha_6 \sim 10^{-2} \text{ K}^{-1}$.

The axial variation of the dimensionless electrothermal force with the thermal perturbation is shown in Fig. 2a. Under isothermal condition $(\xi=0)$, the electric field remains unaffected along the microchannel representing the scenario of purely electroosmotic flow. As we increase the value of $\xi$, an electrothermal force is induced in the axial direction thus altering the temperature distribution significantly. Since the electric field $\left(\bar{E}_x=-\partial\bar{\phi}/\partial\bar{x}\right)$ in the non-isothermal condition depends on the temperature distribution, it gets distorted with increasing $\xi$.

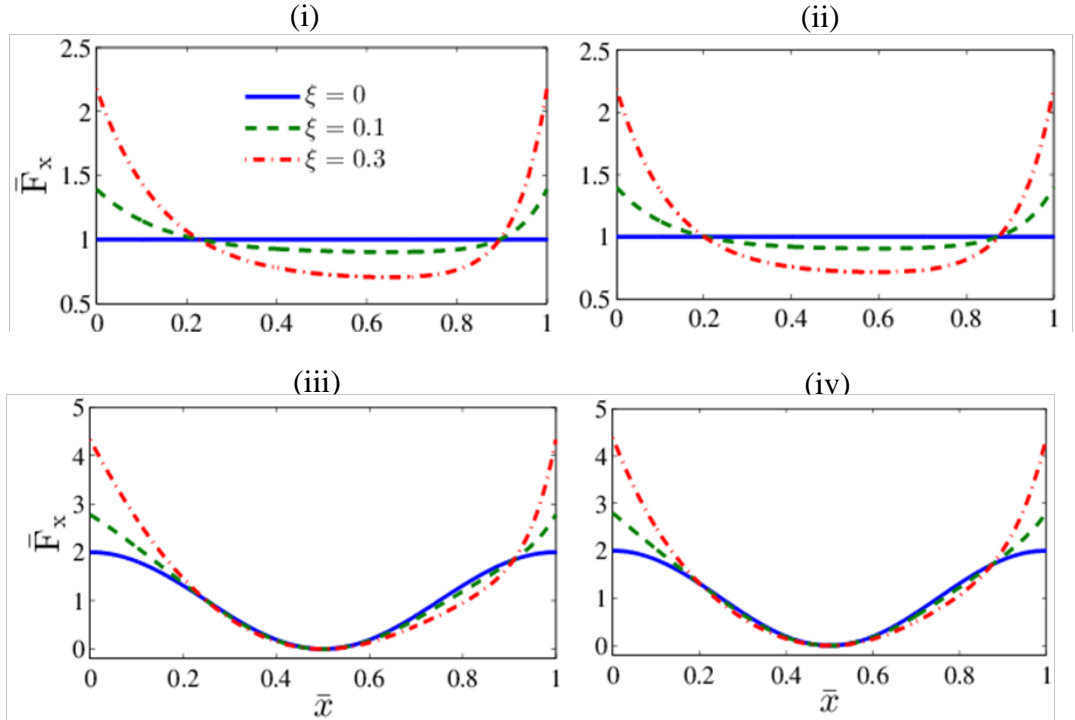

FIG. 2a. The axial variation of the electrothermal force for different values of $\xi$. (i),(iii) Viscoelastic fluid $De=0.5$ and (ii),(iv) Newtonian fluid $De=0$. Also, (i),(ii) represents the case of constant zeta potential $\alpha_2=0$ while (iii),(iv) represents the axially modulated zeta potential $\alpha_2=1$.

When the channel surface is maintained at constant zeta potential (i.e. $\alpha_2=0$), the electrothermal force $\left(\bar{F}_x\right)$ remains constant for isothermal condition, i.e. $\xi=0$. As we move into the non-isothermal scenario, $\bar{F}_x$ starts to deviate from its constant value, gets amplified in the region where the temperature gradients are located. The interaction between the temperature gradient and the surface potential gives rise to an enhancement of the electrothermal force up to two times at the channel ends (at $\xi = 0.3$), clearly seen from Fig. 2a. In absence of any thermal perturbation, the axially modulated potentials result in a harmonic distribution of the electrothermal force with the minima shifted from $\bar{x} = 0.65$ to $\bar{x} = 0.5$. Nevertheless, as we increase the value of $\xi$, the non-trivial interaction between the modulated wall potential and the Joule heating effect affects the imposed temperature gradients at both ends, thereby resulting in an irregular distribution of $\bar{F}_x$.

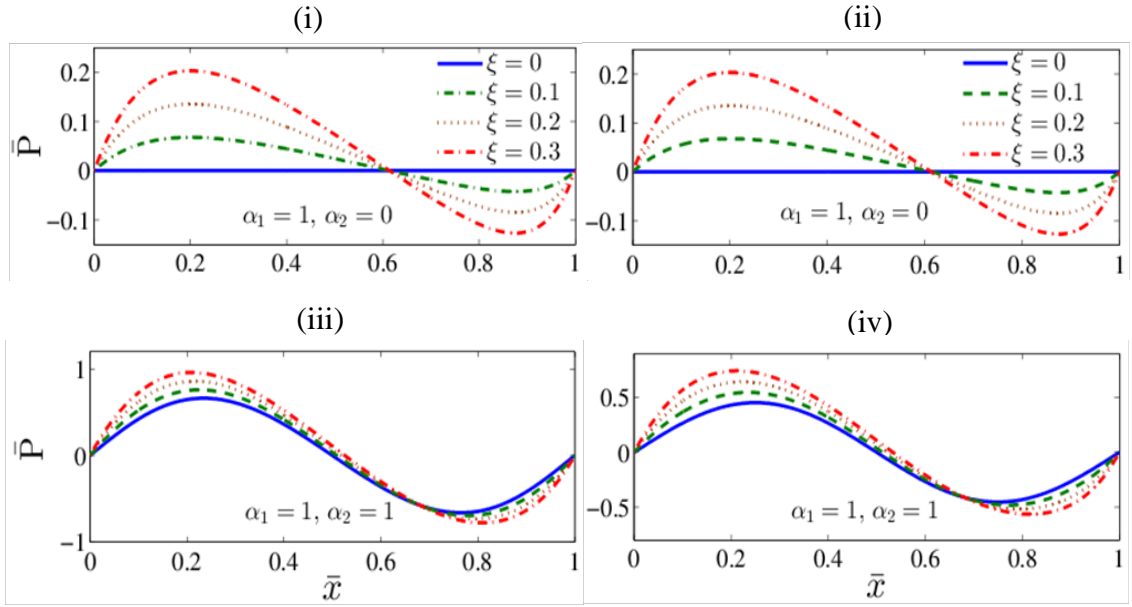

FIG. 2b. Distribution of pressure in the *x*-direction for different $\alpha_2$. For (i),(ii) $\alpha_2 = 0$; for (iii),(iv) $\alpha_2 = 1$. Also (i),(iii) accounts for viscoelastic fluids while (ii),(iv) shows its Newtonian counterpart.

The pressure distribution along the *x*-direction is demonstrated in Fig. 2b. For constant zeta potential, the pressure distribution remains identical for Newtonian fluid and its viscoelastic counterparts. With increasing $\xi$, the interaction between the thermal gradients and the electrokinetic force leads to an imbalance in the pressure distribution, creating over pressure towards the inlet and low pressure at the outlets which in turn affects the electrothermal force. In contrast, the non-homogeneous interaction between the thermal gradients and the modulated potential strongly influences the pressure distribution which is further strengthened in viscoelastic fluids, as can be seen from Fig. 2b. However, the thermal perturbation is less pronounced in viscoelastic fluid. For example, with higher degree of thermal perturbation, the increment of pressure is about ~ 65% for Newtonian fluid while it is ~ 45% for viscoelastic fluids.

The enhancement in the volumetric flow rate is manifested in Fig. 3a where $\bar{Q}$ is plotted as a function of the patterning frequency. The net throughput basically depends on the interaction between the modulated electrokinetic forces and the temperature gradients. This interaction makes the distribution of the flow rate periodic in nature where the degree of periodicity depends on the direction of the electrothermal force and the temperature gradients, dictated by the patterning frequency of modulation. Fig. 3a also includes the variation of $\bar{Q}$ with the increasing Deborah number $(De)$. Increasing the value of $De$ is associated with the pronounced shear-thinning behavior which leads to significant augmentation in the flow rate. It increases linearly up to certain value of Deborah number, $De \sim 0.05$. Then it changes its linearity and increases abruptly following an exponential behavior. Now, the imposed non-isothermal condition induces electrothermal interaction and results in the reduction of the viscosity of the fluid thus leading to the increment of the flow rate as a consequence. The prediction in the thin EDL limit is also incorporated which shows that for thin EDL, the non-linear behavior is observed earlier (i.e. at low values of $De$).

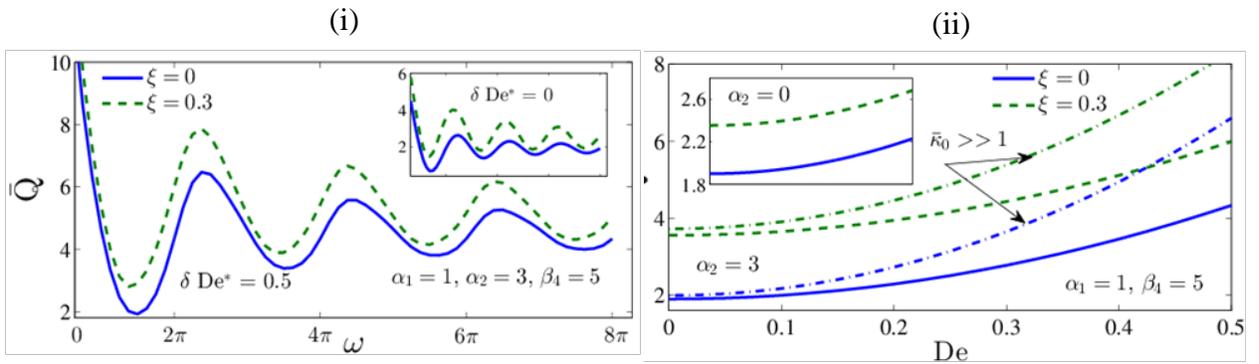

FIG. 3a. (i) The variation of the volumetric flow rate as a function of patterning frequency for $De = 0.5$ (Inset shows the corresponding results for Newtonian fluid). (ii) Dependence of $\bar{Q}$ on Deborah number $(De)$ evaluated at $\alpha_2 = 3$ (Inset for $\alpha_2 = 0$). Also, the predictions in the thin EDL limit are shown by dash-dot lines.

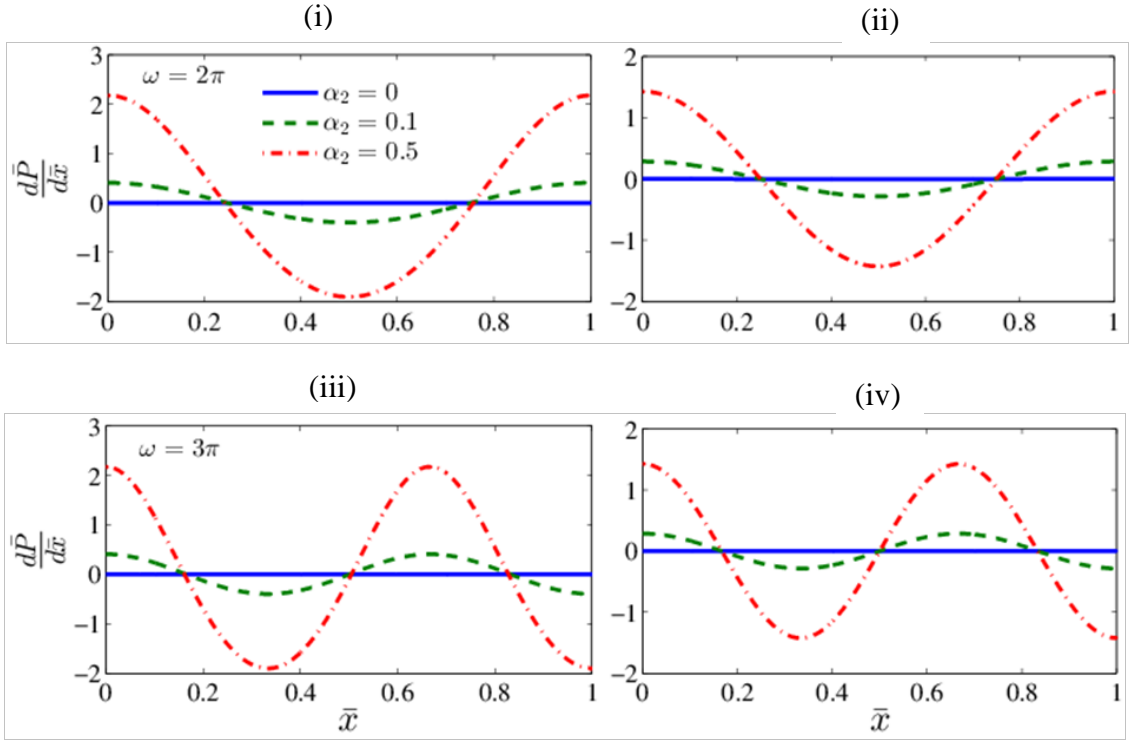

FIG. 3b. The axial variation of the pressure gradient for different patterning frequencies. (i),(ii) corresponds to $\omega = 2\pi$ while (iii),(iv) are for $\omega = 3\pi$. Also (i),(iii) accounts for viscoelastic fluids while (ii),(iv) shows its Newtonian counterpart.

The distribution of the pressure gradient is illustrated in Fig. 3b. No pressure gradient is induced in the *x*-direction when the channel walls are subjected to uniform wall potential (i.e. $\alpha_2 = 0$) and the corresponding velocity profile is plug flow type, typically observed in purely electroosmotic flows. Conversely, increasing the value of $\alpha_2$ strongly influences the hydrodynamics of the flow by creating adverse pressure gradient at both ends and favorable pressure gradient in the middle. In presence of adverse pressure gradient, the flow field is distorted which becomes more pronounced at higher values of $\alpha_2$. On the other hand, favorable pressure gradients results in a convex type of velocity profile and the degree of convexity increases with higher potential modulation $(\alpha_2)$. For patterning frequency $\omega = 2\pi$, the modulated electrokinetic force at the two ends are in the same direction (both positive) while changing the patterning frequency from $\omega = 2\pi$ to $\omega = 3\pi$ makes the forces opposite in the two ends (positive at the inlet and negative at the exit) thus giving rise to two regions of adverse pressure gradients. Also, the region of favorable pressure gradient occurs much earlier due to the imbalance between the forces, as evident from Fig. 3b where the minima is shifted to $\bar{x} = 0.35$ from $\bar{x} = 0.5$ (for the previous case) and the maxima occurs at $\bar{x} = 0.67$, much earlier than the channel exit (which is the location of maxima for $\omega = 2\pi$).

The axial variation of the dispersion coefficient is depicted in Fig. 4a. where the effect of the varying zeta potential is shown. In absence of any modulation in zeta potential (i.e. $\alpha_2 = 0$), it represents purely diffusional dispersion and the corresponding dispersion coefficient is very close to unity $\bar{D}_{eff} \approx 1.0165$, which is enhanced only by $\sim 1\%$ in case of viscoelastic fluid $De = 0.5$. As we start increasing the value of $\alpha_2$, the modulated zeta potential alters the velocity distribution significantly and it no longer remains uniform in the axial direction thus resulting convex and concave profiles which influences strongly the distribution of the dispersion coefficient along the channel. Since adverse pressure gradient prevails up to $\bar{x} = 0.2$, the velocity profile is distorted and follows a concave shape which results in the reduction of the dispersion coefficient and the reduction becomes more prominent at higher modulation $(\alpha_2)$ of surface potential. Increasing the value of $\alpha_2$ induces more non-uniformity in the flow field thereby influencing the velocity gradient strongly. Similarly, favorable pressure gradient is prevalent at the middle of the channel which results in an extra contribution in dispersion in addition to the purely diffusional dispersion. As a result, the dispersion coefficient is enhanced significantly with the maximum increment of $\sim 13.5\%$ observed at $\bar{x} = 0.5$. Besides, the effect of viscoelasticity on the dispersion coefficient is also reflected in Fig 4a. Increasing $De$ is augments the shear-thinning behavior of the fluid resulting in an enhancement of $\bar{D}_{eff}$ up to $\sim 9\%$, as clear from Fig. 4a (v). It is important to mention that, the effect of viscoelasticity is noticeable only in the favorable pressure gradient region and less significant in the adverse pressure gradient region.

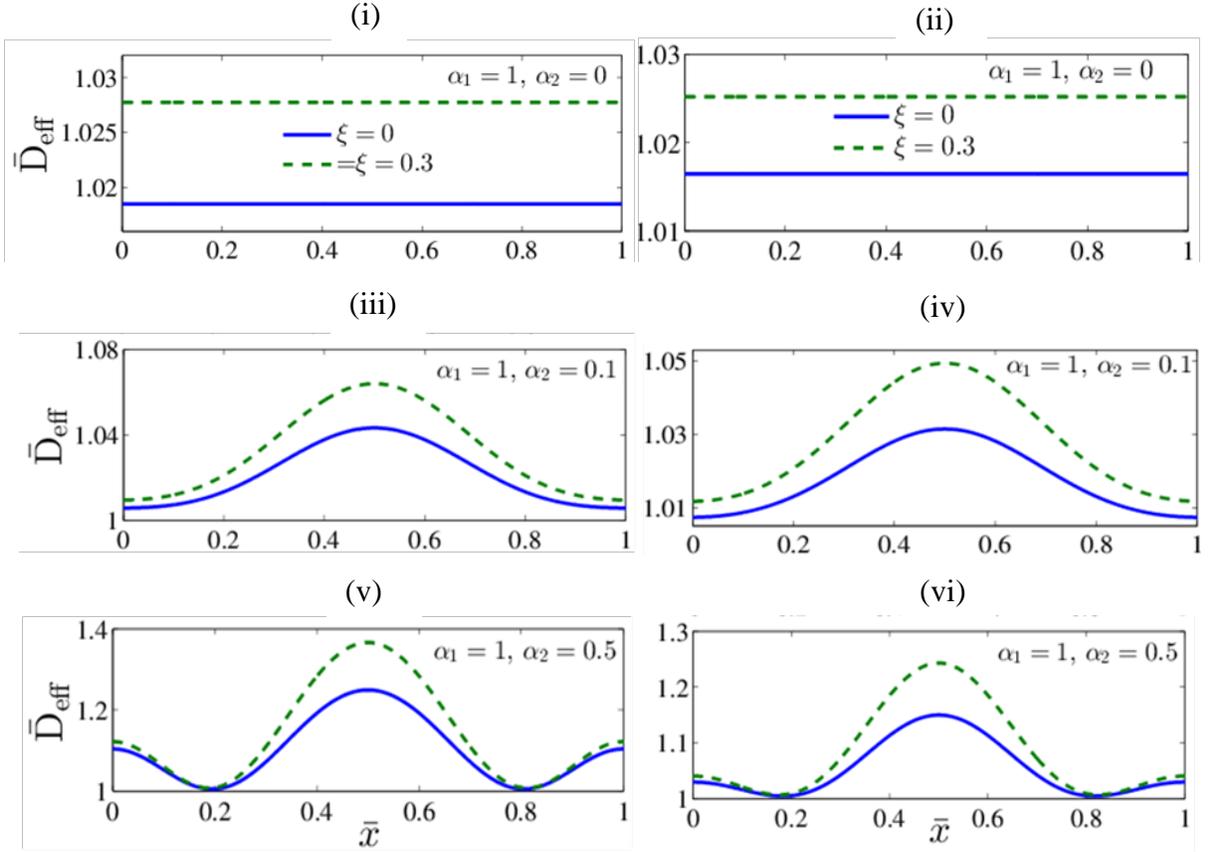

FIG. 4a. The dependence of the dispersion coefficient in the *x*-direction for different values of $\alpha_2$. In (i) and (ii) $\alpha_2 = 0$, in (iii) and (iv) $\alpha_2 = 0.1$, in (v) and (vi) $\alpha_2 = 0.3$. Also, (i),(iii),(v) accounts for viscoelastic fluid $De = 0.5$, (ii),(iv),(vi) for Newtonian fluid.

Now the introduction of the non-isothermal condition induces an electrothermal force which affects the flow field strongly. The presence of temperature gradients at both ends makes the electrothermal force maximum while at the middle, it becomes minimum. So the maximum enhancement of $\bar{D}_{eff}$ is observed at $\bar{x} = 0.5$. Also, increasing the degree of non-uniformity in zeta potential results in the reduction of the electrothermal force and $\bar{D}_{eff}$ is increased up to $\sim 28\%$ by changing the value of $\alpha_2$ from 0.1 to 0.5.

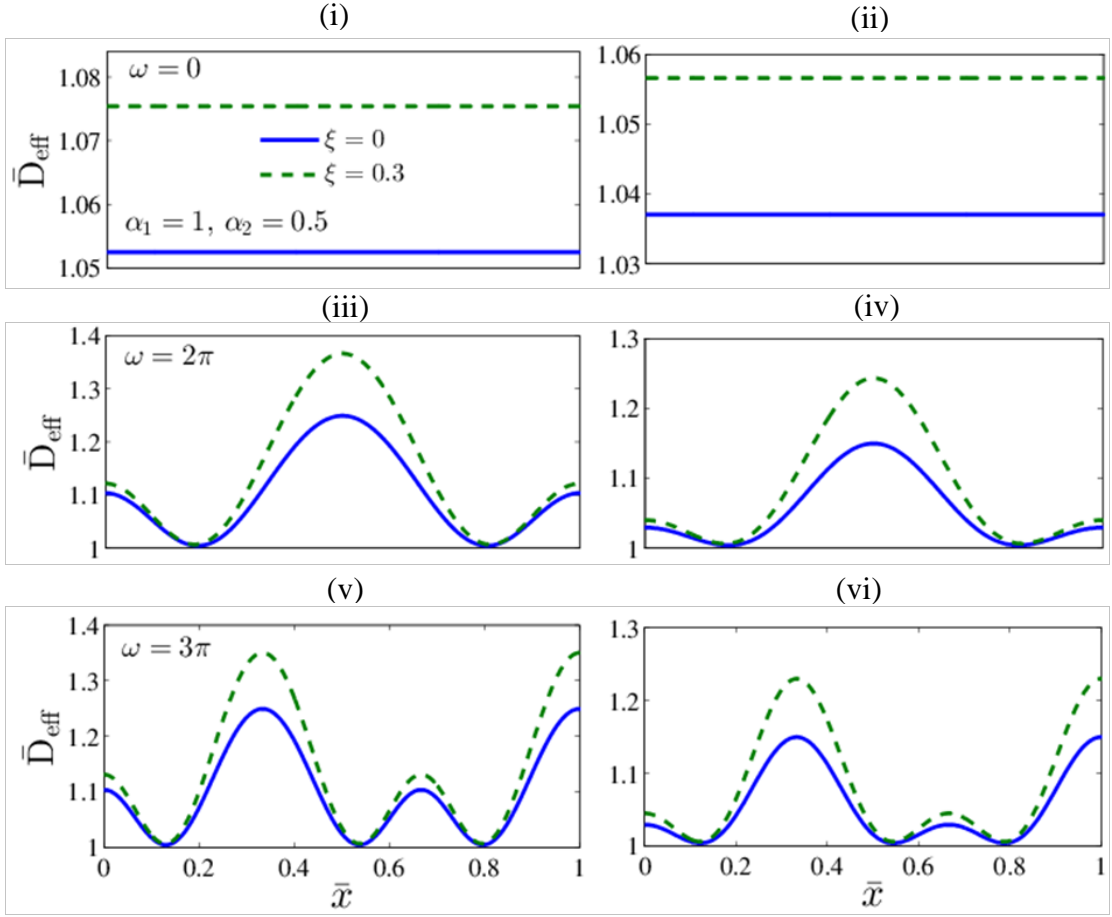

FIG. 4b. The effect of patterning frequency on the axial variation of the dispersion coefficient. Results are shown for two different values of $\xi$. (i),(iii),(v) viscoelastic fluid $De = 0.5$ and (ii), (iv), (vi) Newtonian fluid $De = 0$.

Now the effect of patterning frequency on the dispersion coefficient is highlighted in Fig. 4b. $\omega = 0$ means the axially varying component of the electrokinetic force is absent and the dispersion coefficient remains constant which is further enhanced on imposition of non-isothermal condition $\left(\xi = 0.3\right)$. One interesting thing to observe here that the distribution of $\bar{D}_{eff}$ is altered drastically as we change the patterning frequency from $\omega = 2\pi$ to $\omega = 3\pi$. For $\omega = 3\pi$, the electrokinetic force is modulated in the opposite way and as a result, the region of adverse and favorable pressure gradients is shifted (as shown earlier in Fig. 3b) and two peaks are observed instead of one in the distribution of $\bar{D}_{eff}$. For the sake of brevity, the corresponding results in the thin EDL limit are presented and discussed in **Appendix G**.

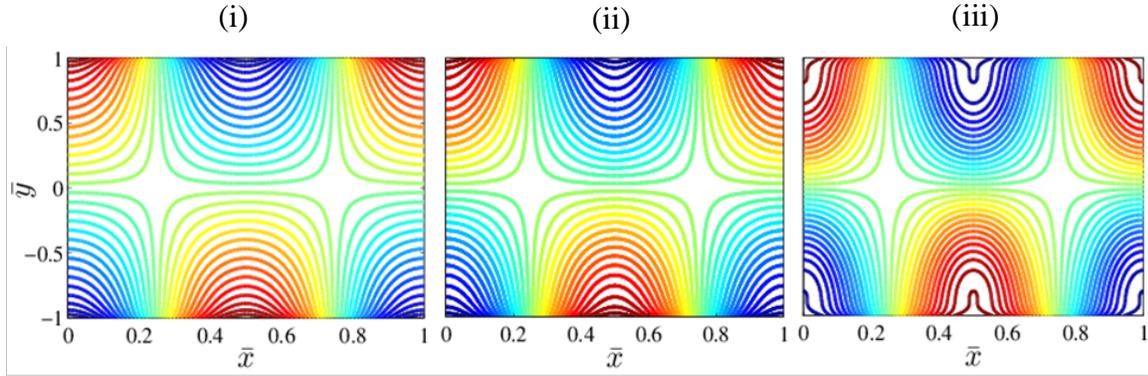

FIG. 5a. Stream function for different values of Deborah number. Results are shown for $\alpha_1 = 0$, $\alpha_2 = 3$ and $\omega = 2\pi$. (i) $De = 0$, (ii) $De = 0.25$ and (iii) $De = 0.5$.

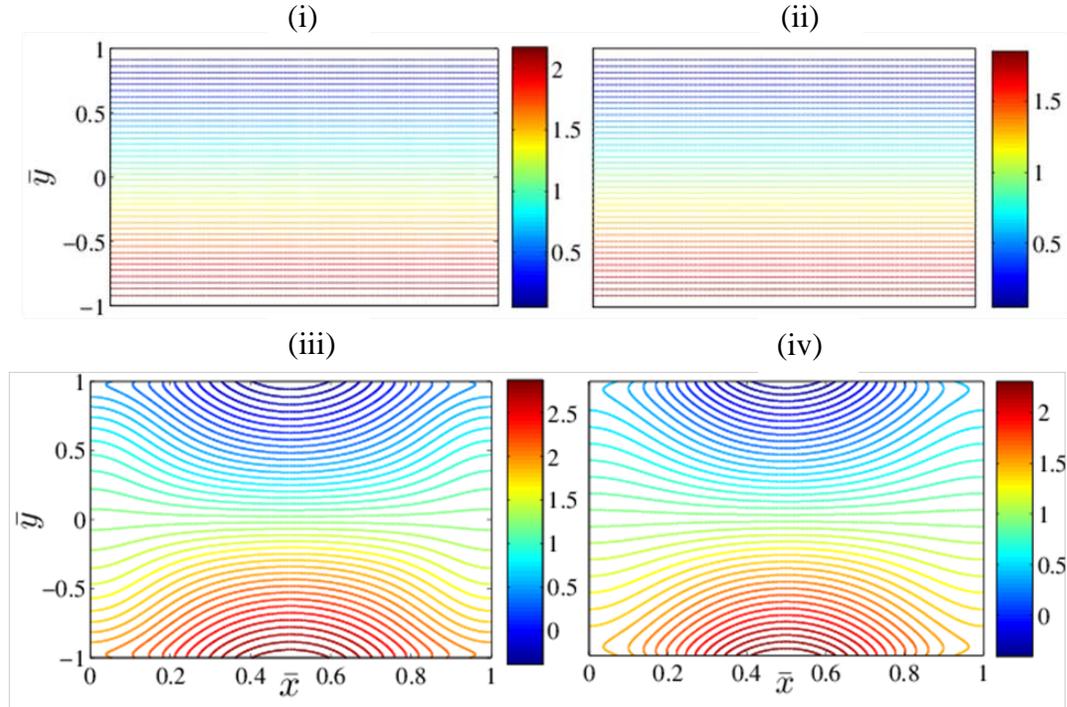

FIG. 5b. The variation of the stream function for different values of $\alpha_2$. (i),(ii) for $\alpha_2 = 0$ and (iii),(iv) for $\alpha_2 = 1$. Also, (i),(iii), for viscoelastic fluid $De = 0.5$ and (ii),(iv), for Newtonian fluid $De = 0$.

The proper interpretation of the flow physics can be done from the streamline contour plots of Figs. 5a-5c where the variation of the stream function is shown. Fig. 5a shows the effect of viscoelasticity on the flow field where the profile for $De = 0$ represents the Newtonian behavior. As we increase the value of Deborah number $(De)$, deviation from Newtonian like behavior is clearly visible and it becomes exaggerated at higher values of $De$. This is attributed to the fact that the stresses are very high near the walls where the effect of EDL on the flow field is observed whereas in the bulk, stresses are relatively lower. Intuitively, the effect is viscoelasticity is observed at its highest level near the wall while very weak influence can be noticed near the channel centreline.

For constant surface potential, the streamlines are straight lines which shows some wavy nature when the potential is axially modulated (Fig. 5b). For low values of patterning frequency $(\omega)$, the invariant component of the potential dictates the flow distribution as shown by the straight lines in Fig. 5c. On the other hand, the axial modulation of the potential starts to the dominate the flow field at relatively higher values of $\omega$. As clear from Fig. 5c, with increasing $\omega$, the undulation of the periodic behavior is more and large recirculation rolls can be observed in the flow which was absent for low values of $\omega$.

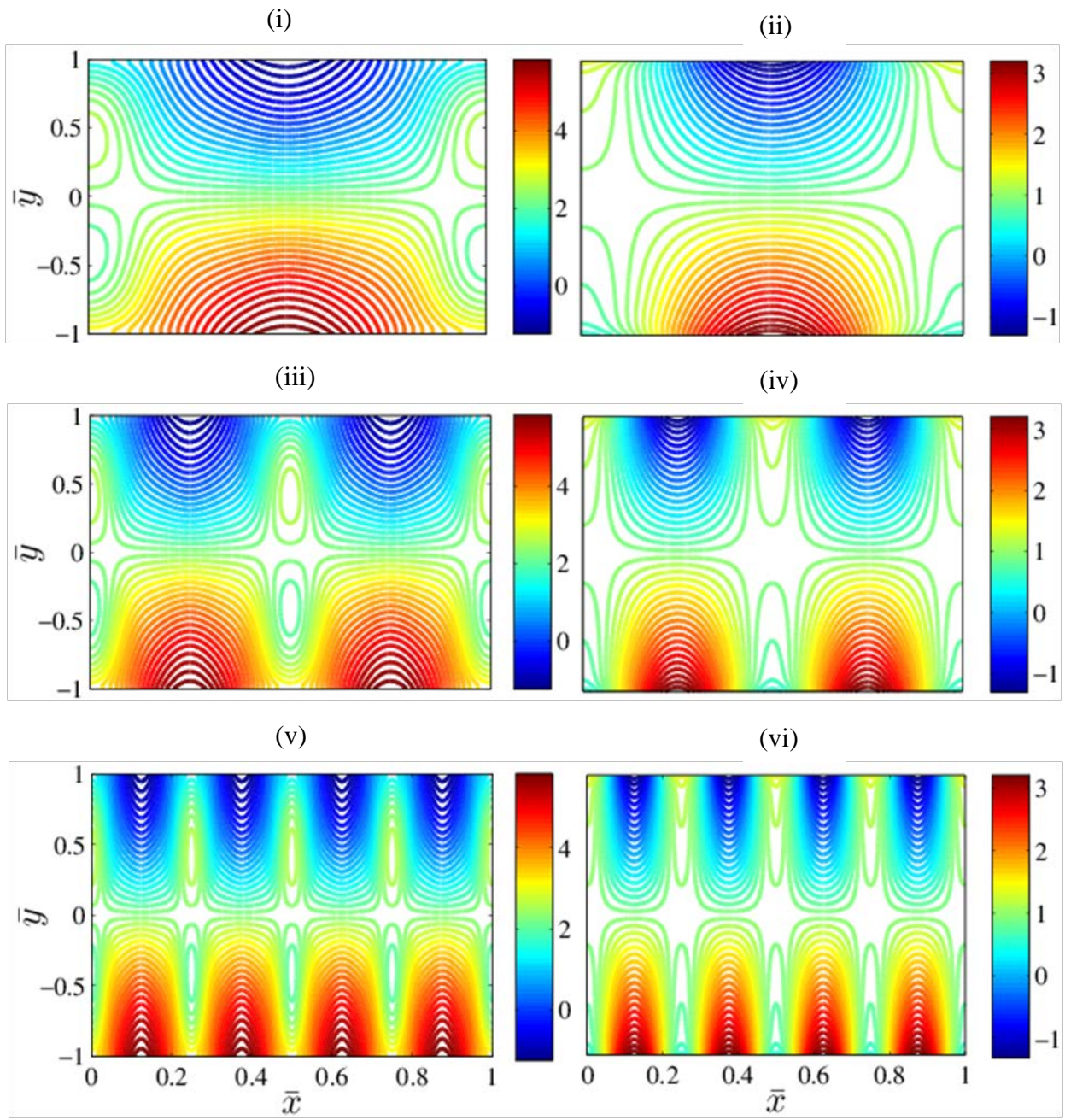

FIG. 5c. The dependence of stream function on $\alpha_2$. (i),(iii),(v) for viscoelastic fluid $De = 0.5$; (ii),(iv),(vi) for Newtonian fluid $De = 0$. Results are reported for (i),(ii) $\omega = 2\pi$, (iii),(iv) $\omega = 4\pi$ and (v),(vi) $\omega = 8\pi$ respectively.

## IV. CONCLUSIONS

We have attempted to delineate the effect of thermal perturbation on patterned electrothermal flow of viscoelastic fluids in a slit microchannel. Considering the physical properties of the fluid to be temperature dependent, the interaction between the temperature gradient and the modulated electrokinetic forces is studied in narrow confinements. To showcase this complex coupling, a regular perturbation approach is taken into consideration after following the classical lubrication approximation theory. First, we bring out the coupling between the modulated surface potential and the temperature gradient to see the effect on the net electrothermal force. Following this, the effect of the electrothermal force of the flow field is investigated which in turn is employed to determine the dispersion coefficient. Since the classical electroosmotic flow results in a uniform velocity profile, it diminishes the extent of mixing. In this context, the present study reveals that, by tuning properly some pertinent parameters like axial modulation, patterning frequency, degree of heating and viscoelasticity one can enhance the dispersion considerably without compromising the volumetric flow rate. Hence, this analysis holds practical relevance in dealing with the electrothermal instabilities associated with the temperature gradient thus can be of utilitarian importance in the management of thermo-fluidic devices as well as in the designing of electrically actuated mixing purposes.

**APPENDIX A: Reasons behind the assumption of Poisson-Boltzmann distribution**

While obtaining the charge distribution, we have assumed that the Poisson-Boltzmann description remains valid. This assumption is based on the fact that ions are point charges and they are in local equilibrium. Under this condition, one can neglect the contribution of the advection term in the Poisson-Nernst-Plank equation when the value of ionic Peclet number $Pe_i$ is very small as compared to unity $(Pe_i \ll 1)$. From definition, $Pe_i$ is written as $Pe_i = u_{ref} h/D$ where $u_{ref}$ is the characteristic velocity scale defined by $u_{ref} \sim \varepsilon_{ref} \zeta_{ref}^2 / \mu_{ref} h$. In typical microfluidic applications, $\varepsilon_{ref} \sim 10^{-10}$ C V$^{-1}$ m$^{-1}$, $\zeta_{ref} \sim 10^{-2}$ V, $\mu_{ref} \sim 10^{-3}$ Pa.s, and $D \sim 10^{-9}$ m$^2$s$^{-1}$ [74] which yields $Pe_i \sim \varepsilon_{ref} \zeta_{ref}^2 / \mu_{ref} D \sim \mathrm{O}(10^{-2})$, for which the effect of ionic mobility can be neglected. In this context, it is also necessary to mention that the Poisson-Boltzmann description of the charge distribution breaks down in presence of finite sized ionic species in which one needs to take into account a more realistic model, commonly termed as modified Poisson-Boltzmann equation. [31,75,76] So, the present analysis is valid only when the effect of ionic mobility and finite size (also known as steric factor) are neglected.

**APPENDIX B: Asymptotic Solution of Eqs. (26)-(27)**

To obtain the asymptotic solution, we have used the well known regular perturbation technique where $\xi$ is chosen as the gauge function to show the effect of thermal perturbation in the flow and temperature distribution. In the limit of $\xi \to 0$, any variable $\gamma$ can be expanded in the following way

$$\gamma = \gamma_0 + \xi\gamma_1 + \xi^2\gamma_2 + \cdots\cdots \tag{40}$$

For leading order, i.e. $O(\xi^0)$, the set of equations are given below

$$\left.\begin{array}{l} \dfrac{\partial \bar{u}_0}{\partial \bar{x}} + \dfrac{\partial \bar{v}_0}{\partial \bar{y}} = 0 \\[6pt] \dfrac{d\bar{p}_0}{d\bar{x}} = \dfrac{\partial \bar{\tau}_{xy,0}}{\partial \bar{y}} - \bar{\kappa}_0^2\{\alpha_1 + \alpha_2\cos(\omega\bar{x})\}\dfrac{\cosh(\bar{\kappa}_0\bar{y})}{\cosh(\bar{\kappa}_0)}\dfrac{d\bar{\phi}_0}{d\bar{x}} \\[6pt] \dfrac{1}{2}Pe\dfrac{d\bar{T}_{0,0}}{d\bar{x}}\left\{\int_{-1}^{1}\bar{u}_0\,d\bar{y}\right\} = \chi\dfrac{d^2\bar{T}_{0,0}}{d\bar{x}^2} - \dfrac{\nu\bar{T}_{0,0}}{\chi} + \left(\dfrac{d\bar{\phi}_0}{d\bar{x}}\right)^2 \\[6pt] \dfrac{d^2\bar{\phi}_0}{d\bar{x}^2} = 0 \end{array}\right\} \tag{41}$$

and
$$\dfrac{\partial \bar{u}_0}{\partial \bar{y}} = \left\{1 + \dfrac{2\delta De_{\bar{\kappa}_0}^2}{\bar{\kappa}_0^2}\bar{\tau}_{yx,0}^2\right\}\bar{\tau}_{yx,0} \tag{42}$$

For first order, i.e. $O(\xi^1)$,

$$\left.\begin{array}{l} \dfrac{\partial \bar{u}_1}{\partial \bar{x}} + \dfrac{\partial \bar{v}_1}{\partial \bar{y}} = 0 \\[6pt] \dfrac{d\bar{p}_1}{d\bar{x}} = \dfrac{\partial \bar{\tau}_{xy,1}}{\partial \bar{y}} - \bar{T}_{0,0}\dfrac{\partial \bar{\tau}_{xy,0}}{\partial \bar{y}} - \bar{\kappa}_0^2\{\alpha_1 + \alpha_2\cos(\omega\bar{x})\}\dfrac{\cosh(\bar{\kappa}_0\bar{y})}{\cosh(\bar{\kappa}_0)}\dfrac{d\bar{\phi}_1}{d\bar{x}} \\[6pt] \beta_4\dfrac{d}{d\bar{x}}\left(\bar{T}_{0,0}\dfrac{d\bar{\phi}_0}{d\bar{x}}\right) + \dfrac{d^2\bar{\phi}_1}{d\bar{x}^2} = 0 \end{array}\right\} \tag{43}$$

$$\text{and} \quad \frac{\partial \bar{u}_1}{\partial \bar{y}} = \bar{\tau}_{yx,1} + \frac{6\delta De_{\bar{\kappa}_0}^2}{\bar{\kappa}_0^2} \bar{\tau}_{yx,0}^2 \bar{\tau}_{yx,1} \tag{44}$$

The boundary conditions described by Eq. (15) are rewritten in the following

$$\left.\begin{aligned}
&\bar{u}_0(\bar{y}=\pm 1) = \bar{u}_1(\bar{y}=\pm 1) = 0, \\
&\bar{v}_0(\bar{y}=\pm 1) = \bar{v}_1(\bar{y}=\pm 1) = 0, \\
&\bar{p}_0(\bar{x}=0,1) = \bar{p}_1(\bar{x}=0,1) = 0, \\
&\bar{T}_{0,0}(\bar{x}=0,1) = \bar{T}_{0,1}(\bar{x}=0,1) = 0, \\
&\bar{\phi}_0(\bar{x}=0) = 1, \ \bar{\phi}_1(\bar{x}=0) = 0, \\
&\bar{\phi}_0(\bar{x}=1) = 0, \ \bar{\phi}_1(\bar{x}=1) = 0.
\end{aligned}\right\} \tag{45}$$

Now, the stress component used in Eq. (42) is substituted by the following expression

$$\bar{\tau}_{xy,0} = \frac{d\bar{p}_0}{d\bar{x}} \bar{y} - \bar{\kappa}_0 \{\alpha_1 + \alpha_2 \cos(\omega \bar{x})\} \frac{\sinh(\bar{\kappa}_0 \bar{y})}{\cosh(\bar{\kappa}_0)} \tag{46}$$

where symmetry condition is taken into account at the channel centreline, i.e. $\bar{\tau}_{xy,0}(\bar{y}=0) = 0$.

Hence, the simplified momentum equation in the leading order takes the form

$$\begin{aligned}
\frac{\partial \bar{u}_0}{\partial \bar{y}} &= \frac{d\bar{p}_0}{d\bar{x}} \bar{y} - \bar{\kappa}_0^2 \{\alpha_1 + \alpha_2 \cos(\omega \bar{x})\} \frac{\sinh(\bar{\kappa}_0 \bar{y})}{\cosh(\bar{\kappa}_0)} \\
&+ \frac{2\delta}{\bar{\kappa}_0^2} De_{\bar{\kappa}_0}^2 \left\{ \frac{d\bar{p}_0}{d\bar{x}} \bar{y} - \bar{\kappa}_0^2 \{\alpha_1 + \alpha_2 \cos(\omega \bar{x})\} \frac{\sinh(\bar{\kappa}_0 \bar{y})}{\cosh(\bar{\kappa}_0)} \right\}^3
\end{aligned} \tag{47}$$

To solve this non-linear equation, one needs to use the same asymptotic approach and the variables are expanded in a similar fashion using $De^*$ as the perturbation parameter defined as $De^* = De^2$. Hence, all $O(1)$ terms represent the Newtonian contribution while $O(De^*)$ and higher order terms are showing the viscoelastic counterpart. Now, the variables are expanded in the following way

$$\left.\begin{aligned}
\bar{u}_0 &= \bar{u}_{0,0} + De^* \bar{u}_{0,1} + De^{*2} \bar{u}_{0,2} + \cdots \\
\bar{v}_0 &= \bar{v}_{0,0} + De^* \bar{v}_{0,1} + De^{*2} \bar{v}_{0,2} + \cdots \\
\bar{p}_0 &= \bar{p}_{0,0} + De^* \bar{p}_{0,1} + De^{*2} \bar{p}_{0,2} + \cdots
\end{aligned}\right\}$$

Now we expand the variables of Eq. (47) and the equations are given below

For $O(1)$:
$$\frac{\partial \bar{u}_{0,0}}{\partial \bar{y}} = \frac{d\bar{p}_{0,0}}{d\bar{x}}\bar{y} - \bar{\kappa}_0^2\{\alpha_1 + \alpha_2 \cos(\omega \bar{x})\}\frac{\sinh(\bar{\kappa}_0 \bar{y})}{\cosh(\bar{\kappa}_0)} \quad (48)$$

For $O(De^*)$:
$$\frac{\partial \bar{u}_{0,1}}{\partial \bar{y}} = \frac{d\bar{p}_{0,1}}{d\bar{x}}\bar{y} + \frac{2\delta}{\bar{\kappa}_0^2}\left[\begin{array}{l}\bar{y}^3\left(\frac{d\bar{p}_{0,0}}{d\bar{x}}\right)^3 - \bar{\kappa}_0^3\frac{\sinh^3(\bar{\kappa}_0 \bar{y})}{\cosh^3(\bar{\kappa}_0)}\{\alpha_1 + \alpha_2 \cos(\omega \bar{x})\}^3 \\ +3\bar{\kappa}_0^2 \bar{y}\frac{\sinh^2(\bar{\kappa}_0 \bar{y})}{\cosh^2(\bar{\kappa}_0)}\{\alpha_1 + \alpha_2 \cos(\omega \bar{x})\}^2\left(\frac{d\bar{p}_{0,0}}{d\bar{x}}\right) \\ -3\bar{\kappa}_0 \bar{y}^2\frac{\sinh(\bar{\kappa}_0 \bar{y})}{\cosh(\bar{\kappa}_0)}\{\alpha_1 + \alpha_2 \cos(\omega \bar{x})\}\left(\frac{d\bar{p}_{0,0}}{d\bar{x}}\right)^2\end{array}\right] \quad (49)$$

Now the solution of Eq. (48) subjected to the no-slip boundary condition is given by

$$\bar{u}_{0,0} = \frac{1}{2}\frac{d\bar{p}_{0,0}}{d\bar{x}}(\bar{y}^2 - 1) + \{\alpha_1 + \alpha_2 \cos(\omega \bar{x})\}\left[1 - \frac{\cosh(\bar{\kappa}_0 \bar{y})}{\cosh(\bar{\kappa}_0)}\right] \quad (50)$$

where the pressure gradient $\frac{d\bar{p}_{0,0}}{d\bar{x}}$ is yet to be determined. This can be done by invoking the continuity equation to determine the *v*-component of the flow field which is then subjected to the impermeability condition at the surfaces $\bar{v}_0(\bar{y} = \pm 1) = 0$ and yields

$$\bar{p}_{0,0} = \frac{3\alpha_2}{\omega}\left[1 - \frac{\tanh(\bar{\kappa}_0)}{\bar{\kappa}_0}\right]\{\sin(\omega \bar{x}) - \bar{x}\sin(\omega)\} \quad (51)$$

and
$$\frac{d\bar{p}_{0,0}}{d\bar{x}} = 3\alpha_2\left[1 - \frac{\tanh(\bar{\kappa}_0)}{\bar{\kappa}_0}\right]\left\{\cos(\omega \bar{x}) - \frac{\sin(\omega)}{\omega}\right\} \quad (52)$$

Similarly proceeding, the solution of Eq. (49) is given by

$$\bar{u}_{0,1} = \frac{1}{2}\frac{d\bar{p}_{0,1}}{d\bar{x}}(\bar{y}^2 - 1) + \frac{2\delta}{\bar{\kappa}_0^2}\left[\begin{array}{l}\frac{a_1^3}{4}\left\{\cos(\omega \bar{x}) - \frac{\sin(\omega)}{\omega}\right\}^3(\bar{y}^4 - 1) - \frac{\bar{\kappa}_0^2\{\alpha_1 + \alpha_2 \cos(\omega \bar{x})\}^3}{12\cosh^3(\bar{\kappa}_0)}f_3(\bar{y}) \\ -\frac{3a_1^2\{\alpha_1 + \alpha_2 \cos(\omega \bar{x})\}}{\bar{\kappa}_0^2 \cosh(\bar{\kappa}_0)}\left\{\cos(\omega \bar{x}) - \frac{\sin(\omega)}{\omega}\right\}^2 f_1(\bar{y}) \\ +\frac{3a_1\{\alpha_1 + \alpha_2 \cos(\omega \bar{x})\}^2}{8\cosh^2(\bar{\kappa}_0)}\left\{\cos(\omega \bar{x}) - \frac{\sin(\omega)}{\omega}\right\}f_2(\bar{y})\end{array}\right] \quad (53)$$

along with the pressure distribution

$$\bar{p}_{0,1} = \frac{3\delta}{\bar{\kappa}_0^2} \begin{bmatrix} -\dfrac{a_1^3}{15\omega^3} f_1(\bar{x}) + \dfrac{a_2 f_1(\bar{\kappa}_0) \alpha_2}{2\bar{\kappa}_0 \omega^3} f_2(\bar{x}) + \dfrac{a_3 f_1(\bar{\kappa}_0)}{18 \bar{\kappa}_0 \omega^2} f_3(\bar{x}) \\ -\dfrac{a_4 f_2(\bar{\kappa}_0)}{27 \bar{\kappa}_0 \alpha_2 \omega} f_4(\bar{x}) - \dfrac{a_5 f_2(\bar{\kappa}_0) \alpha_2}{3 \bar{\kappa}_0 \omega^2} f_5(\bar{x}) - \dfrac{8 a_6 f_3(\bar{\kappa}_0) \alpha_2}{\bar{\kappa}_0 \omega} f_6(\bar{x}) \end{bmatrix} + c_1 \bar{x} \quad (54)$$

The coefficients of Eqs. (51)-(54) are given in **Appendix C**. Once the velocity distribution is known, one can evaluate the corresponding temperature and potential distribution

$$\bar{T}_{0,0} = \dfrac{\dfrac{\chi}{v}\{\exp(d_1)-1\}}{\exp(d_2)-\exp(d_1)}\{\exp(d_2 \bar{x})-\exp(d_1 \bar{x})\} + \dfrac{\chi}{v}\{1-\exp(d_1 \bar{x})\} \quad (55)$$

$$\text{and} \quad \bar{\phi}_0 = 1 - \bar{x} \quad (56)$$

The coefficients of Eq. (55) can be found in **Appendix D**.

Knowing the leading order temperature and potential distribution, one can calculate the higher order potential distribution, as evident from Eq. (43)

$$\bar{\phi}_1 = \beta_4 \left[ \dfrac{\dfrac{\chi}{v}\{\exp(d_1)-1\}}{\exp(d_2)-\exp(d_1)} \right] \left\{ \dfrac{\exp(d_2 \bar{x}) - \bar{x}\exp(d_2) + \bar{x} - 1}{d_2} + \dfrac{1-\bar{x}}{d_1} + \dfrac{\bar{x}\exp(d_1) - \exp(d_1 \bar{x})}{d_1} \right\}$$

$$+ \dfrac{\beta_4 \chi}{v} \left\{ \dfrac{\bar{x}\exp(d_1) - \exp(d_1 \bar{x}) - \bar{x} + 1}{d_1} \right\} \quad (57)$$

$$\dfrac{d\bar{\phi}_1}{d\bar{x}}\bigg|_1 = \beta_4 \left[ \dfrac{\dfrac{\chi}{v}\{\exp(d_1)-1\}}{\exp(d_2)-\exp(d_1)} \right] \left\{ \dfrac{1-\exp(d_2)}{d_2} + \dfrac{\exp(d_1)-1}{d_1} + \exp(d_2 \bar{x}) - \exp(d_1 \bar{x}) \right\}$$

and

$$+ \dfrac{\beta_4 \chi}{v} \left\{ \dfrac{\exp(d_1)-1}{d_1} - \exp(d_1 \bar{x}) \right\} \quad (58)$$

Once $\bar{\phi}_0$ and $\bar{\phi}_1$ are known, the electrothermal body force $(\bar{F}_x)$ in the momentum equation can be evaluated as

$$\bar{F}_x = -\dfrac{1}{2\bar{\kappa}_0} \int_{-1}^{1} \bar{\kappa}_0^2 \{\alpha_1 + \alpha_2 \cos(\omega \bar{x})\} \dfrac{\cosh(\bar{\kappa}_0 \bar{y})}{\cosh(\bar{\kappa}_0)} \dfrac{\partial \bar{\phi}}{\partial \bar{x}} d\bar{y} \quad (59)$$

The solution for the set of equations of $O(\xi^1)$ described by Eqs. (43)-(44) are presented in **Appendix E**. Once the solution is obtained, volumetric flow rate $\bar{Q}$ through the microchannel can be calculated as

$$\bar{Q} = \int_{-1}^{1} \bar{u} \, d\bar{y} = \int_{-1}^{1} (\bar{u}_0 + \xi \bar{u}_1) \, d\bar{y}$$
$$= \int_{-1}^{1} \left( \bar{u}_{0,0} + De^* \bar{u}_{0,1} + \xi \bar{u}_1 + O(\xi^2) + O(De^{*2}) + O(\xi De^*) + \ldots \right) d\bar{y} \tag{60}$$

where the solution is presented corrected up to the first order $O(\xi^1)$ because of the inherent non-linearity of the governing equations.

**APPENDIX C: The coefficients of Eqs. (51)-(54)**

$$a_1 = 3\alpha_2 \left[ 1 - \frac{\tanh(\bar{\kappa}_0)}{\bar{\kappa}_0} \right], a_2 = a_3 = \frac{3a_1^2}{\bar{\kappa}_0^2 \cosh(\bar{\kappa}_0)}, a_4 = a_5 = \frac{3a_1}{8\cosh^2(\bar{\kappa}_0)}, a_6 = \frac{\bar{\kappa}_0^2}{12\cosh^3(\bar{\kappa}_0)} \tag{C1}$$

$$a_1 = 3\alpha_2 \left[ 1 - \frac{\tanh(\bar{\kappa}_0)}{\bar{\kappa}_0} \right], a_2 = a_3 = \frac{3a_1^2}{\bar{\kappa}_0^2 \cosh(\bar{\kappa}_0)}, a_4 = a_5 = \frac{3a_1}{8\cosh^2(\bar{\kappa}_0)}, a_6 = \frac{\bar{\kappa}_0^2}{12\cosh^3(\bar{\kappa}_0)} \tag{C2}$$

$$f_1(\bar{y}) = \left[ (\bar{\kappa}_0^2 \bar{y}^2 + 2)\cosh(\bar{\kappa}_0 \bar{y}) - 2\bar{\kappa}_0 \bar{y} \sinh(\bar{\kappa}_0 \bar{y}) - (\bar{\kappa}_0^2 + 2)\cosh(\bar{\kappa}_0) + 2\bar{\kappa}_0 \sinh(\bar{\kappa}_0) \right]$$
$$f_2(\bar{y}) = \left[ 2\bar{\kappa}_0 \bar{y} \{\sinh(2\bar{\kappa}_0 \bar{y}) - \bar{\kappa}_0 \bar{y}\} - \cosh(2\bar{\kappa}_0 \bar{y}) - 2\bar{\kappa}_0 \{\sinh(2\bar{\kappa}_0) - \bar{\kappa}_0\} + \cosh(2\bar{\kappa}_0) \right] \tag{C3}$$
$$f_3(\bar{y}) = \left[ \cosh(3\bar{\kappa}_0 \bar{y}) - 9\cosh(\bar{\kappa}_0 \bar{y}) - \cosh(3\bar{\kappa}_0) + 9\cosh(\bar{\kappa}_0) \right]$$

$$f_1(\bar{\kappa}_0) = \left[ 6\bar{\kappa}_0^2 \sinh(\bar{\kappa}_0) - 2\bar{\kappa}_0^3 \cosh(\bar{\kappa}_0) - 12\bar{\kappa}_0 \cosh(\bar{\kappa}_0) + 12\sinh(\bar{\kappa}_0) \right]$$
$$f_2(\bar{\kappa}_0) = \left[ -4\bar{\kappa}_0^3 + 6\bar{\kappa}_0^2 \sinh(2\bar{\kappa}_0) - 6\bar{\kappa}_0 \cosh(2\bar{\kappa}_0) + 3\sinh(2\bar{\kappa}_0) \right] \tag{C4}$$
$$f_3(\bar{\kappa}_0) = \left[ -3\bar{\kappa}_0 \cosh^3(\bar{\kappa}_0) + \sinh(\bar{\kappa}_0)\cosh^2(\bar{\kappa}_0) + 9\bar{\kappa}_0 \cosh(\bar{\kappa}_0) - 7\sinh(\bar{\kappa}_0) \right]$$

$$f_1(\bar{x}) = \left[ \{2\omega^2 \cos^2(\omega \bar{x}) + 4\omega^2 - 9\omega \cos(\omega \bar{x})\sin(\omega) + 18 - 18\cos^2(\omega)\}\sin(\omega \bar{x}) + 9\omega^2 \bar{x} \sin(\omega) \right]$$
$$f_2(\bar{x}) = \left[ \left\{ -\frac{2}{9}\omega^2 \cos^2(\omega \bar{x}) + \omega \sin(\omega)\cos(\omega \bar{x}) - \frac{4}{9}\omega^2 + 2\cos^2(\omega) - 2 \right\} \sin(\omega \bar{x}) \right. \tag{C5}$$
$$\left. + \bar{x} \sin(\omega) \left\{ \omega^2 - \frac{2}{3}\cos^2(\omega) + \frac{2}{3} \right\} \right]$$

$$f_3(\bar{x}) = \left[ \left\{ \begin{array}{l} 4\alpha_2 \omega \cos^2(\omega) + (9A\omega - 9\alpha_2 \sin \omega)\cos(\omega \bar{x}) \\ -36\alpha_1 \sin(\omega) + 8\alpha_2 \omega \end{array} \right\} \sin(\omega \bar{x}) + 9\omega \bar{x} \{\alpha_1 \omega - \alpha_2 \sin(\omega)\} \right]$$
(C6)

$$f_4(\bar{x}) = \left[ \left\{ 2\alpha_2^3 \cos^2(\omega \bar{x}) + 9\alpha_1 \alpha_2^2 \cos(\omega \bar{x}) + 18\alpha_1^2 \alpha_2 + 4\alpha_2^3 \right\} \sin(\omega \bar{x}) + 6\alpha_1 \omega \left( \alpha_1^2 + \frac{3}{2}\alpha_2^2 \right) \bar{x} \right]$$

$$f_5(\bar{x}) = \left[ \left\{ \begin{array}{l} \frac{4}{9}\alpha_2 \omega \cos^2(\omega \bar{x}) + (\alpha_1 \omega - \alpha_2 \sin \omega)\cos(\omega \bar{x}) \\ -4\alpha_1 \sin(\omega) + \frac{8}{9}\alpha_2 \omega \end{array} \right\} \sin(\omega \bar{x}) + \omega \bar{x} \{\alpha_1 \omega - \alpha_2 \sin(\omega)\} \right]$$
(C7)

$$f_6(\bar{x}) = \left[ \left\{ \frac{1}{9}\alpha_2^2 \cos^2(\omega \bar{x}) + \frac{1}{2}\alpha_1 \alpha_2 \cos(\omega \bar{x}) + \alpha_1^2 + \frac{2}{9}\alpha_2^2 \right\} \sin(\omega \bar{x}) - \frac{1}{2}\alpha_1 \alpha_2 \omega \bar{x} \right]$$

$$c_1 = \left( -\frac{3\delta}{\bar{\kappa}_0^2} \right) \left[ \begin{array}{l} -\dfrac{a_1^3}{15\omega^3} f_1(\omega) + \dfrac{a_2 f_1(\bar{\kappa}_0)\alpha_2}{2\bar{\kappa}_0 \omega^3} f_2(\omega) + \dfrac{a_3 f_1(\bar{\kappa}_0)}{18\bar{\kappa}_0 \omega^2} f_3(\omega) \\ -\dfrac{a_4 f_2(\bar{\kappa}_0)}{27\bar{\kappa}_0 \alpha_2 \omega} f_4(\omega) - \dfrac{a_5 f_2(\bar{\kappa}_0)\alpha_2}{3\bar{\kappa}_0 \omega^2} f_5(\omega) - \dfrac{8a_6 f_3(\bar{\kappa}_0)\alpha_2}{\bar{\kappa}_0 \omega} f_6(\omega) \end{array} \right]$$
(C8)

$$f_1(\omega) = \left[ \left\{ (2\omega^2 - 18)\cos^2(\omega) + 4\omega^2 - 9\omega \cos(\omega)\sin(\omega) + 18 \right\} \sin(\omega) + 9\omega^2 \sin(\omega) \right]$$

$$f_2(\omega) = \left[ \left\{ 2\omega^2 \cos^2(\omega) - 9\omega \sin(\omega)\cos(\omega) - 5\omega^2 - 12\cos^2(\omega) + 12 \right\} \sin(\omega) \right]$$

$$f_3(\omega) = \left[ \left\{ -9\alpha_2 \cos(\omega) - 36\alpha_1 \right\} \sin^2(\omega) + 9\omega \sin(\omega) \left\{ \frac{4\alpha_2}{9}\cos^2(\omega) + \alpha_1 \cos(\omega) - \frac{\alpha_2}{9} + 9\alpha_1 \omega^2 \right\} \right]$$

$$f_4(\omega) = \left[ \left\{ 2\alpha_2^3 \cos^2(\omega \bar{x}) + 9\alpha_1 \alpha_2^2 \cos(\omega \bar{x}) + 18\alpha_1^2 \alpha_2 + 4\alpha_2^3 \right\} \sin(\omega \bar{x}) + 6\alpha_1 \omega \left( \alpha_1^2 + \frac{3}{2}\alpha_2^2 \right) \bar{x} \right]$$
(C9)

$$f_5(\omega) = \left[ \left\{ \frac{4}{9}\alpha_2 \cos^2(\omega) + \alpha_1 \cos(\omega) - \frac{\alpha_2}{9} \right\} \omega \sin(\omega) + \alpha_1 \omega^2 + \sin^2(\omega)\{-\alpha_2 \cos(\omega) - 4\alpha_1\} \right]$$

$$f_6(\omega) = \left[ \left\{ \frac{1}{9}\alpha_2^2 \cos^2(\omega) + \frac{1}{2}\alpha_1 \alpha_2 \cos(\omega) + \alpha_1^2 + \frac{2}{9}\alpha_2^2 \right\} \sin(\omega) - \frac{1}{2}\alpha_1 \alpha_2 \omega \right]$$

**APPENDIX D: The coefficients of Eq. (55)**

$$d_1 = \frac{\left[ b_1 Pe_T + \sqrt{b_1^2 Pe_T^2 + 4\nu} \right]}{2\chi}, \quad d_2 = \frac{\left[ b_1 Pe_T - \sqrt{b_1^2 Pe_T^2 + 4\nu} \right]}{2\chi}$$

and $b_1 = \frac{1}{2}\int_{-1}^{1} \bar{u}_0 \, d\bar{y} = \frac{1}{2}\int_{-1}^{1}(\bar{u}_{0,0} + De^*\bar{u}_{0,1}) \, d\bar{y}$; where $2b_1$ represents the leading order volumetric flow rate through the microchannel i.e., in absence of any thermal perturbation.

**APPENDIX E: The solution of Eqs. (43)-(44)**

First, we have obtained the stress component of the first order which in turn is used to determine the flow field

$$\bar{\tau}_{xy,1} = \frac{d\bar{p}_1}{dx}\bar{y} + \bar{T}_{0,0}\frac{d\bar{p}_0}{dx}\bar{y} + \bar{\kappa}_0\{\alpha_1 + \alpha_2\cos(\omega\bar{x})\}\frac{\sinh(\bar{\kappa}_0\bar{y})}{\cosh(\bar{\kappa}_0)}\frac{d\bar{\phi}_1}{dx}$$
$$-\bar{T}_{0,0}\bar{\kappa}_0\{\alpha_1 + \alpha_2\cos(\omega\bar{x})\}\frac{\sinh(\bar{\kappa}_0\bar{y})}{\cosh(\bar{\kappa}_0)} \tag{61}$$

$$\frac{\partial \bar{u}_1}{\partial \bar{y}} = \frac{d\bar{p}_1}{dx}\bar{y} + \bar{T}_{0,0}\frac{d\bar{p}_0}{dx}\bar{y} + \bar{\kappa}_0\{\alpha_1 + \alpha_2\cos(\omega\bar{x})\}\frac{\sinh(\bar{\kappa}_0\bar{y})}{\cosh(\bar{\kappa}_0)}\frac{d\bar{\phi}_1}{dx}$$
$$-\bar{T}_{0,0}\bar{\kappa}_0\{\alpha_1 + \alpha_2\cos(\omega\bar{x})\}\frac{\sinh(\bar{\kappa}_0\bar{y})}{\cosh(\bar{\kappa}_0)}$$
$$+\frac{6\delta De_{\bar{\kappa}_0}^2}{\bar{\kappa}_0^2}\left[\frac{d\bar{p}_0}{dx}\bar{y} - \bar{\kappa}_0\{\alpha_1 + \alpha_2\cos(\omega\bar{x})\}\frac{\sinh(\bar{\kappa}_0\bar{y})}{\cosh(\bar{\kappa}_0)}\right]^2 \begin{bmatrix} \frac{d\bar{p}_1}{dx}\bar{y} + \bar{T}_{0,0}\frac{d\bar{p}_0}{dx}\bar{y} \\ +\bar{\kappa}_0\{\alpha_1 + \alpha_2\cos(\omega\bar{x})\}\frac{\sinh(\bar{\kappa}_0\bar{y})}{\cosh(\bar{\kappa}_0)}\frac{d\bar{\phi}_1}{dx} \\ -\bar{T}_{0,0}\bar{\kappa}_0\{\alpha_1 + \alpha_2\cos(\omega\bar{x})\}\frac{\sinh(\bar{\kappa}_0\bar{y})}{\cosh(\bar{\kappa}_0)} \end{bmatrix} \tag{62}$$

To solve this Eq. (62), we have again used the same approach and the variables are expanded as follows

$$\left.\begin{aligned}\bar{u}_1 &= \bar{u}_{1,0} + De^*\bar{u}_{1,1} + De^{*2}\bar{u}_{1,2} + \cdots\cdots \\ \bar{v}_1 &= \bar{v}_{1,0} + De^*\bar{v}_{1,1} + De^{*2}\bar{v}_{1,2} + \cdots\cdots \\ \bar{p}_1 &= \bar{p}_{1,0} + De^*\bar{p}_{1,1} + De^{*2}\bar{p}_{1,2} + \cdots \\ \bar{T}_{0,0} &= \bar{T}_{0,0,0} + De^*\bar{T}_{0,0,1} + De^{*2}\bar{T}_{0,0,2} + \cdots \\ \bar{\phi}_1 &= \bar{\phi}_{1,0} + De^*\bar{\phi}_{1,1} + De^{*2}\bar{\phi}_{1,2} + \cdots\end{aligned}\right\}$$

Thus, Eq. (62) is splitted into two following set of equations

$$\frac{\partial \bar{u}_{1,0}}{\partial \bar{y}} = \frac{d\bar{p}_{1,0}}{d\bar{x}} \bar{y} + \bar{T}_{0,0,0} \frac{d\bar{p}_{0,0}}{d\bar{x}} \bar{y} + \bar{\kappa}_0 \{\alpha_1 + \alpha_2 \cos(\omega \bar{x})\} \frac{\sinh(\bar{\kappa}_0 \bar{y})}{\cosh(\bar{\kappa}_0)} \frac{d\bar{\phi}_{1,0}}{d\bar{x}}$$
$$- \bar{T}_{0,0,0} \bar{\kappa}_0 \{\alpha_1 + \alpha_2 \cos(\omega \bar{x})\} \frac{\sinh(\bar{\kappa}_0 \bar{y})}{\cosh(\bar{\kappa}_0)} \quad (63)$$

$$\frac{\partial \bar{u}_{1,1}}{\partial \bar{y}} = \frac{d\bar{p}_{1,1}}{d\bar{x}} \bar{y} + \bar{T}_{0,0,0} \bar{y} \frac{d\bar{p}_{0,1}}{d\bar{x}} + \bar{T}_{0,0,1} \bar{y} \frac{d\bar{p}_{0,0}}{d\bar{x}}$$
$$+ \bar{\kappa}_0 \{\alpha_1 + \alpha_2 \cos(\omega \bar{x})\} \frac{\sinh(\bar{\kappa}_0 \bar{y})}{\cosh(\bar{\kappa}_0)} \frac{d\bar{\phi}_{1,1}}{d\bar{x}} - \bar{T}_{0,0,1} \bar{\kappa}_0 \{\alpha_1 + \alpha_2 \cos(\omega \bar{x})\} \frac{\sinh(\bar{\kappa}_0 \bar{y})}{\cosh(\bar{\kappa}_0)}$$
$$+ \frac{6\delta}{\bar{\kappa}_0^2} \left[ \frac{d\bar{p}_{0,0}}{d\bar{x}} \bar{y} - \bar{\kappa}_0 \{\alpha_1 + \alpha_2 \cos(\omega \bar{x})\} \frac{\sinh(\bar{\kappa}_0 \bar{y})}{\cosh(\bar{\kappa}_0)} \right]^2 \left[ \begin{array}{c} \frac{d\bar{p}_{1,0}}{d\bar{x}} \bar{y} + \bar{T}_{0,0,0} \frac{d\bar{p}_{0,0}}{d\bar{x}} \bar{y} + \\ \bar{\kappa}_0 \{\alpha_1 + \alpha_2 \cos(\omega \bar{x})\} \frac{\sinh(\bar{\kappa}_0 \bar{y})}{\cosh(\bar{\kappa}_0)} \frac{d\bar{\phi}_{1,0}}{d\bar{x}} \\ - \bar{T}_{0,0,0} \bar{\kappa}_0 \{\alpha_1 + \alpha_2 \cos(\omega \bar{x})\} \frac{\sinh(\bar{\kappa}_0 \bar{y})}{\cosh(\bar{\kappa}_0)} \end{array} \right] \quad (64)$$

The solution of Eq. (63) is given by

$$\bar{u}_{1,0} = \frac{1}{2} \frac{d\bar{p}_{1,0}}{d\bar{x}} (\bar{y}^2 - 1) + \frac{1}{2} \bar{T}_{0,0,0} \frac{d\bar{p}_{0,0}}{d\bar{x}} (\bar{y}^2 - 1) + \{\alpha_1 + \alpha_2 \cos(\omega \bar{x})\} \left\{ \frac{\cosh(\bar{\kappa}_0 \bar{y})}{\cosh(\bar{\kappa}_0)} - 1 \right\} \frac{d\bar{\phi}_{1,0}}{d\bar{x}}$$
$$+ \bar{T}_{0,0,0} \{\alpha_1 + \alpha_2 \cos(\omega \bar{x})\} \left\{ 1 - \frac{\cosh(\bar{\kappa}_0 \bar{y})}{\cosh(\bar{\kappa}_0)} \right\} \quad (65)$$

where $\frac{d\bar{p}_{1,0}}{d\bar{x}}$ is obtained in a similar fashion as mentioned earlier. Finally the expression for the velocity profile can be given by

$$\bar{u} = \bar{u}_{0,0} + De^* \bar{u}_{0,1} + De^{*2} \bar{u}_{0,2} + \xi(\bar{u}_{1,0} + De^* \bar{u}_{1,1}) + \ldots$$
$$= \bar{u}_{0,0} + De^* \bar{u}_{0,1} + \xi \bar{u}_1 + O(\xi^2) + O(De^{*2}) + O(\xi De^*) + \ldots \quad (66)$$

Here, the results are reported correct up to first order where the contributions from the higher order terms are omitted for simplification. The expressions of $\bar{T}_{0,0,0}$ and $\frac{d\bar{\phi}_{1,0}}{d\bar{x}}$ presented in **Appendix F** for the completeness of the problem.

**APPENDIX F: The expressions of $\bar{T}_{0,0,0}$ and $\frac{d\bar{\phi}_{1,0}}{d\bar{x}}$**

$$\bar{T}_{0,0,0} = \frac{\chi}{v} \cdot \frac{\left(\exp(d_{1,0})-1\right)}{\left(\exp(d_{2,0})-\exp(d_{1,0})\right)} \cdot \left(\exp(d_{2,0}\,x)-\exp(d_{1,0}\,x)\right) + \frac{\chi}{v} \cdot \left(1-\exp(d_{1,0}\,x)\right) \quad \text{(F1)}$$

$$\frac{d\bar{\phi}_{1,0}}{d\bar{x}} = \beta_4 \left[ \frac{\frac{\chi}{v}\{\exp(d_{1,0})-1\}}{\exp(d_{2,0})-\exp(d_{1,0})} \right] \left\{ \frac{1-\exp(d_{2,0})}{d_{2,0}} + \frac{\exp(d_{1,0})-1}{d_{1,0}} + \exp(d_{2,0}\,\bar{x}) - \exp(d_{1,0}\,\bar{x}) \right\}$$
$$+ \frac{\beta_4 \chi}{v} \left\{ \frac{\exp(d_{1,0})-1}{d_{1,0}} - \exp(d_{1,0}\,\bar{x}) \right\} \quad \text{(F2)}$$

where $d_{1,0} = \dfrac{\left[b_{1,0}\,Pe_T + \sqrt{b_{1,0}^2\,Pe_T^2 + 4v}\right]}{2\chi}$, $d_{2,0} = \dfrac{\left[b_{1,0}\,Pe_T - \sqrt{b_{1,0}^2\,Pe_T^2 + 4v}\right]}{2\chi}$ and $b_{1,0} = \dfrac{1}{2}\int_{-1}^{1} \bar{u}_{0,0}\, d\bar{y}$

Here, $2b_{1,0}$ represents the volumetric flow rate for patterned electroosmotic flow of a Newtonian fluid given by

$$2b_{1,0} = 2\left[1 - \frac{\tanh(\bar{\kappa}_0)}{\bar{\kappa}_0}\right]\left\{\alpha_1 + \alpha_2 \frac{\sin(\omega)}{\omega}\right\} \quad \text{(F3)}$$

**APPENDIX G: Results in the thin EDL limit**

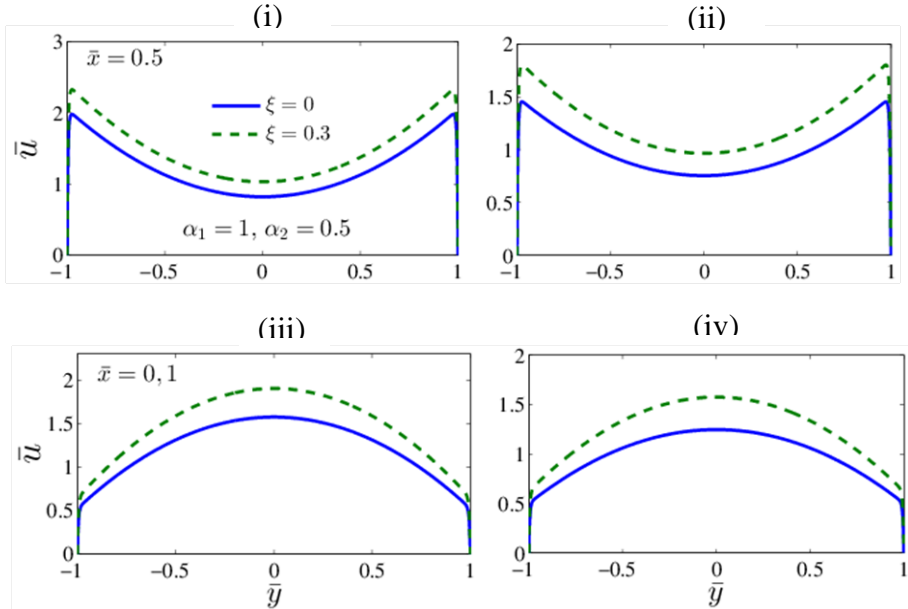

FIG. 6a. Velocity profile in the $y$-direction, evaluated at $\alpha_1 = 1$, $\alpha_2 = 0.5$, $\omega = 2\pi$. (i), (iii) viscoelastic fluid $(De = 0.5)$ and (ii), (iv) Newtonian fluid $(De = 0)$.

When the thickness of the EDL becomes very small (of the order of the few nanometers), the region of excess charge distribution is very less as compared to the channel dimension. In presence of axially modulated surface potential, the favorable pressure gradient for $\omega = 2\pi$ occurs at the middle of the channel while adverse pressure appears to be present at the channel ends the effect of which is clearly reflected in the velocity distribution of Fig. 6a. Similarly, the distribution of the dispersion coefficient shows bimodal behavior with maximum augmentation occurring in the middle and minimum at the two ends. Also, the degree of fluctuation gets amplified on imposition of non-isothermal condition $(\xi)$ as well as fluid viscoelasticity $(De)$. (as shown in Fig. 6b).

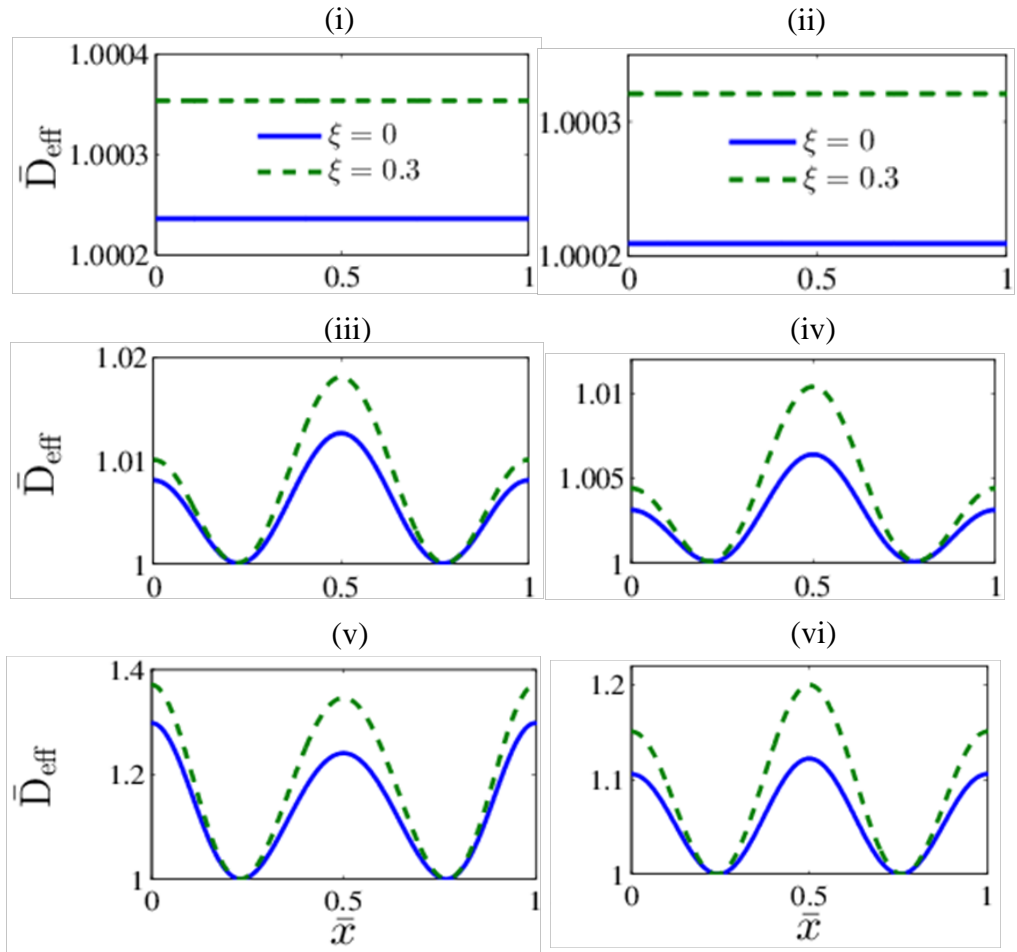

FIG. 6b. The effect of $\omega$ on the axial variation of $\bar{D}_{\mathit{eff}}$, evaluated for different values of $\xi$. (i), (iii) viscoelastic fluid $(De = 0.5)$ and (ii), (iv) Newtonian fluid $(De = 0)$.

**REFERENCES**

[1] G.M. Whitesides, "The origins and the future of microfluidics," Nature, **442**, 368 (2006).

[2] H.A. Stone, A.D. Stroock, and A. Ajdari, "Engineering Flows in Small Devices: Microfluidics Toward a Lab-on-a-Chip," Annu. Rev. Fluid Mech., **36**, 381 (2004).

[3] A.D. Stroock, S.K.W. Dertinger, A. Ajdari, I. Mezic, H.A. Stone, and G.M. Whitesides, "Chaotic mixer for microchannels.," Science, **295**, 647 (2002).

[4] C.C. Chang and R.J. Yang, "Chaotic mixing in a microchannel utilizing periodically switching electro-osmotic recirculating rolls," Phys. Rev. E, **77**, 1 (2008).

[5] H. Sugioka, "Chaotic mixer using electro-osmosis at finite Péclet number," Phys. Rev. E - Stat. Nonlinear, Soft Matter Phys., **81**, 1 (2010).

[6] J.B. Zhang, G.W. He, and F. Liu, "Electro-osmotic flow and mixing in heterogeneous microchannels," Phys. Rev. E, **73**, 1 (2006).

[7] I. Glasgow, J. Batton, and N. Aubry, "Electroosmotic mixing in microchannels.," Lab Chip, **4**, 558 (2004).

[8] P.D. Anderson, O.S. Galaktionov, G.W.M. Peters, F.N. van de Vosse, and H.E.H. Meijer, "Mixing of non-Newtonian fluids in time-periodic cavity flows," J. Nonnewton. Fluid Mech., **93**, 265 (2000).

[9] R.J. Hunter, *Zeta Potential in Colloid Science* (Academic Press, London, 1981).

[10] R. F. Probstein, *Physicochemical Hydrodynamics: An Introduction*, 2nd ed. (Wiley, New York, 1994).

[11] B.A. and A.N. Karniadakis G., *Microflows and Nanoflows* (Springer, 2005).

[12] R. Aris, "On the Dispersion of a Solute in a Fluid Flowing Through a Tube," Proc. R. Soc. A, **235**, 67 (1956).

[13] R. Aris, "On the dispersion of a solute by diffusion, convection, and exchange between phases," Proc. R. Soc. A, **252**, 538 (1959).

[14] E.K. Zholkovskij, J.H. Masliyah, and J. Czarnecki, "Electroosmotic Dispersion in Microchannels with a Thin Double Layer," Anal. Chem., **75**, 901 (2003).

[15] E.K. Zholkovskij and J.H. Masliyah, "Hydrodynamic Dispersion due to Combined Pressure-Driven and Electroosmotic Flow Through Microchannels with a Thin Double Layer," Anal. Chem., **76**, 2708 (2004).

[16] A. Ajdari, N. Bontoux, and H.A. Stone, "Hydrodynamic Dispersion in Shallow Microchannels: the Effect of Cross-Sectional Shape," Anal. Chem., **78**, 387 (2006).

[17] D. Dutta, "Electrokinetic Transport of Charged Samples through Rectangular Channels with Small Zeta Potentials," Anal. Chem., **80**, 4723 (2008).

[18] C.L.A. Berli and M.L. Olivares, "Electrokinetic flow of non-Newtonian fluids in microchannels," J. Colloid Interface Sci., **320**, 582 (2008).

[19] C.L.A. Berli, "Electrokinetic energy conversion in microchannels using polymer solutions," J. Colloid Interface Sci., **349**, 446 (2010).

[20] S. Das and S. Chakraborty, "Analytical solutions for velocity, temperature and concentration distribution in electroosmotic microchannel flows of a non-Newtonian bio-fluid," Anal. Chim. Acta, **559**, 15 (2006).

[21] M.L. Olivares, L. Vera-Candioti, and C.L.A. Berli, "The EOF of polymer solutions," Electrophoresis, **30**, 921 (2009).

[22] C. Zhao and C. Yang, "Electro-osmotic mobility of non-Newtonian fluids," Biomicrofluidics,

**5**, (2011).

[23] C. Zhao and C. Yang, "Electrokinetics of non-Newtonian fluids: A review," Adv. Colloid Interface Sci., **201–202**, 94 (2013).

[24] R.G. Owens, "A new microstructure-based constitutive model for human blood," J. Nonnewton. Fluid Mech., **140**, 57 (2006).

[25] M. Moyers-Gonzalez, R.G. Owens, and J. Fang, "A non-homogeneous constitutive model for human blood. Part 1. Model derivation and steady flow," J. Fluid Mech., **617**, 327 (2008).

[26] A. Vissink, H.A. Waterman, E.J. 's-Gravenmade, A.K. Panders, and A. Vermey, "Rheological properties of saliva substitutes containing mucin, carboxymethylcellulose or polyethylenoxide," J. Oral Pathol. Med., **13**, 22 (1984).

[27] H. Fam, J.T. Bryant, and M. Kontopoulou, "Rheological properties of synovial fluids.," Biorheology, **44**, 59 (2007).

[28] A.F. Silva, M.A. Alves, and M.S.N. Oliveira, "Rheological behaviour of vitreous humour," Rheol. Acta, **56**, 377 (2017).

[29] A.M. Afonso, M.A. Alves, and F.T. Pinho, "Analytical solution of mixed electro-osmotic/pressure driven flows of viscoelastic fluids in microchannels," J. Nonnewton. Fluid Mech., **159**, 50 (2009).

[30] U. Ghosh and S. Chakraborty, "Electroosmosis of viscoelastic fluids over charge modulated surfaces in narrow confinements," Phys. Fluids, **27**, (2015).

[31] S. Mukherjee, P. Goswami, J. Dhar, S. Dasgupta, and S. Chakraborty, "Ion-size dependent electroosmosis of viscoelastic fluids in microfluidic channels with interfacial slip," Phys. Fluids, **29**, 72002 (2017).

[32] L.L. Ferrás, A.M. Afonso, M.A. Alves, J.M. Nóbrega, and F.T. Pinho, "Electro-osmotic and pressure-driven flow of viscoelastic fluids in microchannels: Analytical and semi-analytical solutions," Phys. Fluids, **28**, (2016).

[33] U. Ghosh, K. Chaudhury, and S. Chakraborty, "Electroosmosis over non-uniformly charged surfaces: modified Smoluchowski slip velocity for second-order fluids," J. Fluid Mech., **809**, 664 (2016).

[34] S. Mukherjee, S.S. Das, J. Dhar, S. Chakraborty, and S. DasGupta, "Electroosmosis of Viscoelastic Fluids: Role of Wall Depletion Layer," Langmuir, **33**, 12046 (2017).

[35] S. Haeberle and R. Zengerle, "Microfluidic platforms for lab-on-a-chip applications," Lab Chip, **7**, 1094 (2007).

[36] D. Mark, S. Haeberle, G. Roth, F. von Stetten, and R. Zengerle, "Microfluidic lab-on-a-chip platforms: requirements, characteristics and applications," Chem. Soc. Rev., **39**, 1153 (2010).

[37] M. Brust, C. Schaefer, R. Doerr, L. Pan, M. Garcia, P.E. Arratia, and C. Wagner, "Rheology of Human Blood Plasma : Viscoelastic Versus Newtonian Behavior," Phys. Rev. Lett., **78305**, 6 (2013).

[38] T. Das and S. Chakraborty, "Perspective : Flicking with flow : Can microfluidics revolutionize the cancer research ?," Biomicrofluidics, **11811**, (2014).

[39] H.A. Stone, A.D. Stroock, and A. Ajdari, "Engineering Flows in Small Devices," Annu. Rev. Fluid Mech., **36**, 381 (2004).

[40] A. Ajdari, "Electro-Osmosis on Inhomogeneously Charged Surfaces," Phys. Rev. Lett., **75**, 755 (1995).

[41] A. Ajdari, "Generation of transverse fluid currents and forces by an electric field: Electro-osmosis on charge-modulated and undulated surfaces," Phys. Rev. E, **53**, 4996 (1996).

[42] S. Mandal, U. Ghosh, A. Bandopadhyay, and S. Chakraborty, "Electro-osmosis of superimposed fluids in the presence of modulated charged surfaces in narrow confinements," J. Fluid Mech., **776**, 390 (2015).

[43] U. Ghosh and S. Chakraborty, "Patterned-wettability-induced alteration of electro-osmosis over charge-modulated surfaces in narrow confinements," Phys. Rev. E, **85**, 1 (2012).

[44] W.Y. Ng, S. Goh, Y.C. Lam, C. Yang, and I. Rodríguez, "DC-biased AC-electroosmotic and AC-electrothermal flow mixing in microchannels," Lab Chip, **9**, 802 (2009).

[45] R. V. Craster and O.K. Matar, "Electrically induced pattern formation in thin leaky dielectric films," Phys. Fluids, **17**, (2005).

[46] A. Ajdari, "Pumping liquids using asymmetric electrode arrays," Phys. Rev. E, **61**, R45 (2000).

[47] A. González, A. Ramos, N.G. Green, A. Castellanos, and H. Morgan, "Fluid flow induced by nonuniform ac electric fields in electrolytes on microelectrodes. II. A linear double-layer analysis," Phys. Rev. E, **61**, 4019 (2000).

[48] A. Ramos, A. González, A. Castellanos, N.G. Green, and H. Morgan, "Pumping of liquids with ac voltages applied to asymmetric pairs of microelectrodes," Phys. Rev. E, **67**, 11 (2003).

[49] T.M. Squires and M.Z. Bazant, "Induced-charge electro-osmosis," J. Fluid Mech., **509**, 217 (2004).

[50] O. Schnitzer and E. Yariv, "Induced-charge electro-osmosis beyond weak fields," Phys. Rev. E, **86**, 1 (2012).

[51] A.L. Garcia, L.K. Ista, D.N. Petsev, M.J. O'Brien, P. Bisong, A.A. Mammoli, S.R.J. Brueck, and G.P. López, "Electrokinetic molecular separation in nanoscale fluidic channels," Lab Chip, **5**, 1271 (2005).

[52] S. Ghosal, "Fluid mechanics of electroosmotic flow and its effect on band broadening in capillary electrophoresis," Electrophoresis, **25**, 214 (2004).

[53] A. Bandopadhyay and S. Chakraborty, "Giant augmentations in electro-hydro-dynamic energy conversion efficiencies of nanofluidic devices using viscoelastic fluids," Appl. Phys. Lett., **101**, (2012).

[54] T. Nguyen, Y. Xie, L.J. de Vreede, A. van den Berg, and J.C.T. Eijkel, "Highly enhanced energy conversion from the streaming current by polymer addition," Lab Chip, **13**, 3210 (2013).

[55] C.H. Chen, H. Lin, S.K. Lele, and J.G. Santiago, "Convective and absolute electrokinetic instability with conductivity gradients," J. Fluid Mech., **524**, 263 (2005).

[56] J.R. Melcher, "Traveling-Wave Bulk Electroconvection Induced across a Temperature Gradient," Phys. Fluids, **10**, 1178 (1967).

[57] K.W. Bong, J. Xu, J.H. Kim, S.C. Chapin, M.S. Strano, K.K. Gleason, and P.S. Doyle, "Non-polydimethylsiloxane devices for oxygen-free flow lithography," Nat. Commun., **3**, 805 (2012).

[58] X.C. Xuan, B. Xu, D. Sinton, and D.Q. Li, "Electroosmotic flow with Joule heating effects," Lab Chip, **4**, 230 (2004).

[59] H.C. Feldman, M. Sigurdson, and C.D. Meinhart, "AC electrothermal enhancement of heterogeneous assays in microfluidics," Lab Chip, **7**, 1553 (2007).

[60] H. Zhao, "Streaming potential generated by a pressure-driven flow over superhydrophobic stripes," Phys. Fluids, **23**, 22003 (2011).

[61] N.A. Mortensen, L.H. Olesen, L. Belmon, and H. Bruus, "Electrohydrodynamics of binary electrolytes driven by modulated surface potentials," Phys. Rev. E, **71**, 56306 (2005).

[62] C. Ng and C.Y. Wang, "Oscillatory Flow Through a Channel With Stick-Slip Walls: Complex

Navier's Slip Length," J. Fluids Eng., **133**, 14502 (2011).

[63] D.A. Saville, "ELECTROHYDRODYNAMICS:The Taylor-Melcher Leaky Dielectric Model," Annu. Rev. Fluid Mech., **29**, 27 (1997).

[64] S.B. J. Masliyah, *Electrokinetic and Colloid Transport Phenomena* (Wiley-Interscience, 2006).

[65] X. Xuan, "Joule heating in electrokinetic flow," Electrophoresis, **29**, 33 (2008).

[66] J.R. Melcher and G.I. Taylor, "Electrohydrodynamics: A Review of the Role of Interfacial Shear Stresses," Annu. Rev. Fluid Mech., **1**, 111 (1969).

[67] N.P. Thien and R.I. Tanner, "A new constitutive equation derived from network theory," J. Nonnewton. Fluid Mech., **2**, 353 (1977).

[68] N. P. Thien, "A Nonlinear Network Viscoelastic Model," J. Rheol. (N. Y. N. Y)., **22**, 259 (1978).

[69] M. M. Denn, *Polymer Melt Processing* (Cambridge University Press, 2008).

[70] R. I. Tanner, *Engineering Rheology* (Oxford University Press, 2002).

[71] S. Ghosal, "Lubrication theory for electro-osmotic flow in a microfluidic channel of slowly varying cross-section and wall charge," J. Fluid Mech., **459**, 103 (2002).

[72] O. Bautista, S. Sánchez, J.C. Arcos, and F. Méndez, "Lubrication theory for electro-osmotic flow in a slit microchannel with the Phan-Thien and Tanner model," J. Fluid Mech., **722**, 496 (2013).

[73] J.C. Arcos, F. Méndez, E.G. Bautista, and O. Bautista, "Dispersion coefficient in an electro-osmotic flow of a viscoelastic fluid through a microchannel with a slowly varying wall zeta potential," J. Fluid Mech., **839**, 348 (2018).

[74] M.Z. Bazant, K. Thornton, and A. Ajdari, "Diffuse-charge dynamics in electrochemical systems," Phys. Rev. E, **70**, 21506 (2004).

[75] I. Borukhov, D. Andelman, and H. Orland, "Steric Effects in Electrolytes: A Modified Poisson-Boltzmann Equation," Phys. Rev. Lett., **79**, 435 (1997).

[76] M.S. Kilic, M.Z. Bazant, and A. Ajdari, "Steric effects in the dynamics of electrolytes at large applied voltages. I. Double-layer charging," Phys. Rev. E, **75**, 21502 (2007).